\documentclass[12pt,preprintnumbers,amsmath,amssymb,showpacs]{revtex4}

\usepackage{dcolumn}
\usepackage[dvips]{epsfig}
\usepackage{psfrag}
\usepackage{latexsym}
\usepackage{graphicx}

\newcommand{\be}{\begin{equation}}
\newcommand{\ee}{\end{equation}}
\newcommand{\bea}{\begin{eqnarray}}
\newcommand{\eea}{\end{eqnarray}}
\newcommand{\scrs}{\scriptscriptstyle}
\def\bogo{{Bogomol'nyi}}

\def\Phidag{\Phi^\dagger}
\def\pa{\partial}
\def\[{\left[}
\def\]{\right]}
\def\half{{\mathchoice{{\textstyle{1\over 2}}}{1\over 2}{1\over 2}{1 \over 2}}}
\newcommand{\scr}{\scriptstyle}

\begin{document}

\title{Twisted Superconducting Semilocal Strings}
\author{P\'eter Forg\'acs$^{1,2}$, S\'{e}bastien Reuillon$^{1}$ and  Mikhail S.~Volkov$^{1}$}
 \affiliation{ {$^{1}$Laboratoire de Math\'{e}matiques et Physique Th\'{e}orique\\
CNRS-UMR 6083, Universit\'{e} de Tours,\\
Parc de Grandmont 37200 Tours, France}\\
 {$^{2}$MTA RMKI, H-1525 Budapest, P.O.Box 49, Hungary}
}

\begin{abstract}
A new class of twisted, current carrying,
{\sl stationary}, straight string solutions
having finite energy per unit length is constructed numerically in an
extended Abelian Higgs model with global SU(2) symmetry.
The new solutions correspond to
deformations of the embedded Abrikosov-Nielsen-Olesen
(ANO) vortices by a twist -- a relative coordinate dependent phase between the two Higgs fields.
The twist induces a global current flowing through the string, and the deformed
solutions bifurcate with the ANO vortices in the limit of vanishing current.
For each value of the winding
number $n=1,2\ldots$ (determining the magnetic flux through the plane orthogonal
to the string) there are $n$ distinct, two-parametric families of solutions.
One of the continuously varying parameters is
the twist, or the corresponding current,
the other one can be chosen to be the momentum of the string.
For fixed values of the momentum and twist, the $n$ distinct solutions have
different energies and can be viewed as a lowest energy ``fundamental'' string
and its $n-1$ ``excitations'' characterized by different values of their
``polarization''. The latter is defined as the ratio of the
angular momentum of the vortex and its momentum.
In their rest frame the twisted vortices have
{\sl lower energy} than the embedded ANO vortices and could be of considerable
importance in various physical systems (from condensed matter to cosmic strings).
\end{abstract}
\maketitle

\newpage

\section{Introduction}
It was almost 50 years ago that Abrikosov predicted the existence of vortex
lines \cite{Abrikosov} in type II superconductors in the framework
of the Ginzburg-Landau (GL) theory, which has then
been confirmed  many times by experiments in various superconducting materials.
The importance of such vortex lines or strings in high energy physics
has been recognized by Nielsen and Olesen, who
have shown the existence of classical solutions
describing straight, infinitely long strings with finite energy per
unit length in a relativistic field theory -- in the Abelian Higgs model \cite{NO}.
Ever since their discovery, vortices have been the subject of intense studies,
since they find numerous applications in many domains of physics including
various branches of condensed matter theory (such as superconductivity, superfluid
$^3$He \cite{Salomaa}), high energy physics and cosmology.
In a more general context vortex lines belong to the class
of objects known as ``topological defects''. Topological defects (monopoles,
vortices, domain walls, etc.) result from the spontaneous breaking
of global or local symmetries of the underlying field theory;
we refer to  some of the excellent reviews \cite{review}.
There is an interesting class of models
in which both global and local symmetries are simultaneously broken,
in the most economical way, with the minimal number of scalar fields.
These models, dubbed semilocal models, have some remarkable features;
for a detailed review see Ref.\ \cite{semilocal-review}.
Perhaps the most interesting point is that semilocal models exhibit stable vortices,
despite the first homotopy group of the vacuum manifold being trivial
\cite{vac-ach,hin}.

A particularly interesting and simple class of such
semilocal models is the extended Abelian Higgs model (EAH), which contains
electromagnetism coupled to N
charged complex scalars transforming as a vector in the fundamental representation
of the global SU(N) symmetry group.
The simplest nontrivial EAH model is the SU(2) symmetric (N=2)
one, which  case is made quite interesting by noting
that it represents the bosonic sector of the Electroweak theory in the limit
where the Weinberg angle, $\theta_{\rm W}=\pi/2$ \cite{vac92}.
A non-relativistic version of this model is also
used in condensed matter theory,
in the description of unconventional superconductors for example
\cite{Sigrist}.
The Abrikosov-Nielsen-Olesen (ANO) vortex solutions \cite{NO},
characterized by an integer winding number, $n$, determining their magnetic
flux, and by
the mass ratio, $\sqrt{\beta}=m_{\rm s}/m_{\rm v}$
($m_{\rm s}$ resp.\ $m_{\rm v}$
denoting the mass of the scalar resp.\ vector fields),
can be embedded into this theory.
The parameter $\beta$ also distinguishes superconductors of type I ($\beta<1$)
from type II ($\beta>1$).
The case $\beta=1$ is quite special, for here,
instead of there being a unique vortex solution satisfying
the first order \bogo\ equations \cite{bog},
there is a continuous family of them having the same energy  \cite{hin}.
The stability of the  ANO vortices  embedded into the EAH models was examined in
Ref.\ \cite{hin}, and it was shown there that they are stable only if
$\beta\leq1$.
For $\beta>1$ the embedded ANO vortices have been found to exhibit
a single unstable mode that can be interpreted as ``magnetic spreading''
instability \cite{hin, Preskill} .
The Standard Model counterparts of the embedded ANO solutions are the Z-strings,
which have been shown to be stable for
$\sin^2\theta_{\rm W}\gtrsim0.9$ \cite{instab}.

In this paper we give a detailed description of a new class of twisted,
straight string solutions in the simplest nontrivial SU(2) symmetric
semilocal EAH model,
some of our main results having been announced in Ref.\ \cite{FRV}.
The new solutions are translationally symmetric in the $z=x^3$ direction.
The main feature distinguishing them from the previously known string or
equivalently vortex (when the string is viewed in the $(x^1, x^2)$ plane)
solutions is due to the presence of a relative phase (twist) between the two scalars,
depending linearly on $z$.
The twist induces a {\sl persistent global current} flowing along the $z$-direction.
It turns out that there is a maximal value of the twist, depending on $\beta$.
For this maximal twist
the current vanishes, and the twisted solutions {\sl bifurcate}
with the embedded ANO vortex.
In fact the above mentioned ``magnetic spreading'' instability of the embedded
ANO vortex in the EAH model
can be viewed as a signal of bifurcation with a new branch of current
carrying solutions.

For any fixed value of $\beta>1$, and for a given winding number, $n$,
the static twisted vortices form $n$ distinct one parameter families,
labeled by the twist (or equivalently by the current)
and by an integer $\nu=1,\ldots n$.
For a fixed value of the twist, the $n$ different solutions have different energies and
can be viewed as a lowest energy ``fundamental'' string
and its $n-1$ ``excitations''.
It is quite interesting that the energy of the fundamental twisted solutions is
{\sl lower} than that of the corresponding ANO ones. They could therefore
 have better stability properties, and in fact
we expect the fundamental $n=1$ twisted string
to be stable in the N$=2$ EAH model for any value of $\beta>1$.

In the simplest case the twisted vortices are static.
But the relative phase of the two scalar fields
may also have a linear dependence on the time coordinate, $t=x^0$,
in which case the vortices become {\sl stationary}.
These solutions can be
obtained from the static ones by Lorentz boosts in the $t,z$ plane.
The stationary strings are surrounded by an exponentially screened
radial electric field,
and they also possess a nonzero {\sl momentum},
as well as a nonzero {\sl angular momentum}
oriented in the $z$-direction.
For any fixed value of $\beta>1$, and for a given winding number, $n$,
the twisted, stationary vortices form $n$ distinct two-parameter families
which can be labeled by their momentum and current.
The discrete parameter, $\nu=1,\ldots n$, which distinguishes between the $n$
different families can be interpreted as a kind of ``polarization'',
since it determines the ratio of the momentum and the angular momentum
of the string.
Due to the Lorentz contraction,
the energy of the stationary strings is higher than that
of the static ones.

Our solutions can also be viewed as superconducting cosmic strings
in a ``minimal'' model, with a single U(1) gauge field,
unlike in the prototype U(1)$\times$U(1) model of
superconducting strings introduced by  Witten \cite{Witten}.
In contradistinction to the superconducting strings in Witten's model,
the twisted semilocal strings are strongly localized in the transverse direction,
which guarantees the finiteness of their energy. In addition, their
superconducting current is global and not local, and it can be arbitrarily large,
unlike the situation in Witten's model \cite{current}.

Most previously known semilocal string solutions are static and untwisted,
with the notable exception of the solutions of  Abraham \cite{Abraham},
which have however infinite energy per unit length (for $n=1$).
We should also like to mention that in Ref.\ \cite{Torokoff}, in a
theory similar to the EAH model, straight, twisted string solutions have
been obtained in the description of neutral two-component plasmas.
Finally it should be pointed out that the non-abelian counterparts of
our twisted strings should exist in the Electroweak
model in the form of ``twisted Z-strings'' in analogy to the ordinary
Z-strings.

The plan of the paper is the following: In Section II we introduce the EAH models.
In Section III
the stationary, $z$-translationally symmetric Ansatz is given, and the EAH theory is
dimensionally reduced to two dimensions corresponding to the
plane orthogonal to the symmetry axis. Section IV is devoted to the classification
of the stationary and twisted string configurations according to their properties under
Lorentz transformations. In Section V the theory is further simplified by
considering rotationally
symmetric configurations in the plane. We discuss the field equations,
describe the local behaviour of their solutions in the neighborhood of the
 singular points.
Section VI contains a short review of some relevant known vortex solutions: the
classical ANO vortices, the ``skyrmion'' solutions  satisfying the
\bogo\ equations,
and the chiral, charged vortices of Ref.\ \cite{Abraham}.
Section VII is the main part of the paper, it contains a detailed description of the new
solutions. Some concluding remarks are given in  Section VIII.

\section{Extended Abelian Higgs model}
\setcounter{equation}{0}

The extended Abelian Higgs (EAH) model considered below
is a theory of N complex scalars,
$\Phi^T=(\phi_1,\ldots\phi_{\rm N})$,
with an SU(N) global symmetry and with their overall phase gauged.
Such a theory possesses an SU(N)$_{\rm\scr global}\times$U(1)$_{\rm\scr local}$
symmetry algebra by construction.
In this paper we shall concentrate on the simplest
case corresponding to N=2, mostly because of its relationship to
the electroweak model.
The $4$ dimensional (4D) action of the theory is given by
\be\label{4D-action}
S=\int\! \left(-\frac14\,F_{\mu\nu}F^{\mu\nu}
+(D_\mu\Phi)^\dagger D^\mu\Phi-\frac{\lambda}{2}\,
(\Phi^\dagger\Phi-\eta^2)^2\right)\,
d^4x\,.
\end{equation}
Here $F_{\mu\nu}=\partial_\mu A_\nu-\partial_\nu A_\mu$
is the gauge field tensor,
$D_\mu\Phi=\partial_\mu\Phi-igA_\mu\Phi$ is the gauge
covariant derivative, $g$ denotes the coupling
(charge), and the  signature of the flat
Minkowskian metric used here is $(+,-,-,-)$.
Due to the form of the potential
the scalar field has a vacuum expectation value, $\eta$,
and the U(1) gauge symmetry is
spontaneously broken. The physical spectrum consists of a
vector particle with mass $m_{\rm v}=
\sqrt{2}g\eta$, of $2({\rm N}-1)$ Nambu-Goldstone bosons, and of a Higgs
scalar with mass $m_{\rm s} = \sqrt{2\lambda}\eta$.

It is convenient to scale the fields and coordinates as:
$\Phi\to\eta\Phi$, $A_\mu\to\eta A_\mu$, $x^\mu\to x^\mu/(g\eta)$,
so the rescaled fields and coordinates are
dimensionless. In terms of the rescaled quantities
 the action (\ref{4D-action}) becomes:
\be\label{4D-action1}
S=\frac{1}{g^2}\int  {\cal L}\,d^4 x
=\frac{1}{g^2}\int\! d^4x\,\left\{-\frac14\,F_{\mu\nu}F^{\mu\nu}
+(D_\mu\Phi)^\dagger D^\mu\Phi-\frac{\beta}{2}\,
(\Phi^\dagger\Phi-1)^2\right\}
\,,
\end{equation}
where now $D_\mu\Phi=\partial_\mu\Phi-iA_\mu\Phi$, and $\beta=\lambda/g^2$.

The fields transform under the U(1) gauge symmetry as
\be\label{gauge}
A_\mu\to A_\mu+\partial_\mu\Lambda(x)\,,\qquad
\Phi\to e^{i\Lambda(x)}\Phi\,,
\end{equation}
while the global SU(N) symmetry acts on the fields in the following way:
\be
A_\mu\to A_\mu\,,\qquad
\Phi\to e^{i(\theta_a\tau^a)}\Phi,
\end{equation}
where $\tau^a$ $(a=1,\dots,{\rm N})$ denote the generators of the SU(N) Lie algebra
in the fundamental representation
(for N=2 we choose $\tau^a$ to be the Pauli matrices),
 and the $\theta_a$ are constant parameters.
The corresponding Noether currents are
\be\label{cur}
j_{\mu}^{\hat a}=-i\{(D_\mu\Phi)^\dagger T^{\hat a}\Phi
-\Phi^\dagger T^{\hat a} D_\mu\Phi\},
\end{equation}
with ${\hat a}=(0,a)$ and $T^{\hat a}=({\mathbf{1}},\tau^a)$ being generators of
SU(N)$\times$U(1).
The conservation of the currents (\ref{cur}) follows
from the Euler-Lagrange equations for the
action \eqref{4D-action}, which can be written as:
\begin{subequations}\label{4d-eqs}
\begin{align}
\pa^\rho F_{\rho\mu}& = -j^0_{\mu}=i\{(D_\mu\Phi)^\dagger\Phi
-\Phi^\dagger D_\mu\Phi\}\,, \\
D_\rho D^\rho\Phi& = -\beta(|\Phi|^2-1)\Phi\,,
\end{align}
\end{subequations}
where $|\Phi|^2=\Phi^\dagger\Phi$.

We shall also need the energy momentum tensor induced by varying the metric
tensor:
\be
T^\mu_{\phantom\mu\nu}=2g^{\mu\sigma}\frac{\pa\mathcal{L}}{\pa g^{\sigma\nu}}
-\delta^\mu_{\phantom\mu\nu}\mathcal{L}\,.
\end{equation}
With ${\cal L}$ given by \eqref{4D-action1} this
 evaluates to
\be
T^\mu_{\phantom\mu\nu}=
-F^{\mu\sigma}F_{\nu\sigma}+(D^\mu\Phi)^\dagger D_\nu\Phi
+(D_\nu\Phi)^\dagger D^\mu\Phi-\delta^\mu_{\phantom\mu\nu}\mathcal{L}\,.
\end{equation}
The energy density reads
\be\label{en}
T^0_{\phantom0 0}=
|D_0\Phi|^2 +
|D_i\Phi|^2 + \frac{\beta}{2} \left(|\Phi|^2- 1\right)^2 +
\frac{1}{2} {\vec E}^2 +
\frac{1}{2} {\vec B}^2\, ,
\ee
where the Latin indices take the values $i,j,k = 1,2,3$,
 $|D_i\Phi|^2=(D_i\Phi)^\dagger D_i\Phi$ (summation over the index $i$ is implied),
the electric and\ magnetic fields, ${\vec E}$ and\ ${\vec B}$, are defined
as $E_i=F_{0i}$ and $B_k=\epsilon_{ijk}F_{ij}/2$, respectively.

\section{Dimensional reduction to 2 dimensions}
\setcounter{equation}{0}

Modulo gauge and global symmetry transformations
the ground state of the theory \eqref{4D-action1} (i.e. the solution of zero energy)
is given by
$A_\mu = 0$, $\Phi = \Phi_0$, where $\Phi_0$ is a constant complex N-vector,
($|\Phi_0|^2=1$).
 Thus for the case considered in this paper,
${\rm N}=2$, the vacuum manifold is a 3-sphere:
\be\label{vacman}
{\cal V}  = \{ \Phi \in{\mathbf C}^2 \ | \ \Phidag \Phi =1\}
\cong S^3 \ .
\end{equation}
As is well known, there are no spherically symmetric, finite energy solutions
of the EAH model, but it admits stringlike solutions \cite{semilocal-review}.
These strings are infinitely long
and their total energy is infinite, nevertheless they can be of physical interest
if their energy per unit length is finite.
In what follows we shall be considering such strings in the simplest case when they
are stationary and translationally invariant along the $z=x^3$ axis. In this case
the most general Ansatz
for the vector field $A_\mu$ and for the two (upper and lower) complex components
$\phi_1$ and $\phi_2$ of the Higgs field $\Phi$ can be written as
\begin{align}\label{ansatz1}
&A_\mu =  (A_{\scr 0}(x_1,x_2)\,,A_{\scr 3}(x_1,x_2)\,,A_{\scr i}(x_1,x_2))\equiv
(A_{\alpha}\,,A_{\scr i}),\quad \alpha=0,3\,,~~~\ i=1,2,\notag\\
&\phi_1 = f_1(x_1,x_2), \quad
\phi_2 =  f_2(x_1,x_2)e^{i(\omega_0 t+\omega_3 z)}\,,
\end{align}
where $f_1$,$f_2$ are complex functions and
$(\omega_0,\omega_3)=\omega_\alpha$, are real parameters. This Ansatz
easily follows from the general results of Ref.\ \cite{FM} applied to the case
of two commuting symmetry transformations.
The most important feature of the Ansatz (\ref{ansatz1}) for the present paper is that
there is a nontrivial, coordinate dependent {\sl relative phase} between the two
components of the SU(2) doublet $\phi_1$ and $\phi_2$.
The change in the phase of the scalar fields induced by infinitesimal space-time
translations in the $t$, $z$ directions is compensated by 
gauge transformations (followed by global SU(2) rotations).
In other words a space-time translation
moves the fields {\sl along gauge orbits}. The relative phases, $\omega_\alpha$,
can be interpreted as a relative rotation, $\omega_0$, resp.\ twist
along the $z$-axis, $\omega_3$,  between the two components of the scalar field $(\phi_1,\phi_2)$.
One should also note that the Ansatz (\ref{ansatz1})
breaks the originally present global SU(2) symmetry to U(1).

The vacuum manifold \eqref{vacman} is a three sphere, which is simply connected
($\pi_1(S^3)=1$), and so there are no topological string solutions. However,
as explained in \cite{semilocal-review}, because of the spontaneous breaking of the U(1)
gauge symmetry
the space of finite energy configurations is {not labeled} by the vacuum manifold
${\cal V}$, but rather by the gauge orbits, which space is not simply connected:
$\pi_1($U(1)$_{\rm\scr local}/1)=\mathbb{Z}$.
This implies that configurations with different
U(1)$_{\rm\scr local}$ winding numbers are separated by infinitely high energy barriers.

For field configurations given by Ansatz (\ref{ansatz1}) the action (\ref{4D-action1})
dimensionally reduces from 4D to 2D, $S=\frac{1}{g^2}\int {\cal S}dx^0 dx^3$, where
\be\label{action-red}
{\cal S} = \eta^2\int d^2x \Bigl\{A_\alpha A^\alpha|\phi_a|^2 +
\omega^\alpha(\omega_\alpha-2A_\alpha)|\phi_2|^2 -|D_i \phi_a|^2 +
 \frac{1}{2}\pa_iA_\alpha\pa_iA^\alpha
- \frac{1}{4}F_{ij}^2
-\frac{\beta}{2}\left(|\phi_a|^2 - 1\right)^2 \Bigr\}\,,
\end{equation}
where $a=1,2$;
the index $\alpha=(0,3)$ is raised by the residual
Minkowskian metric, i.e.\
$A^\alpha A_\alpha= A_0^2-A_3^2$, while the 2D metric of the reduced theory
in the $(x^1,x^2)$ plane is Euclidean;
summation over the indices $a,i,j=1,2$ is understood.

The equations of motion of the reduced theory (\ref{action-red})
in the $(x_1,x_2)$ plane take the form:
\begin{subequations}\label{Gauss-law}
\begin{align}
\triangle A_\alpha&= 2A_\alpha |\phi_a|^2-2\omega_\alpha|\phi_2|^2\,,\\
 \pa_iF_{ij}&= i(\phi_a\overline{D_j \phi}_a - \overline{\phi}_a D_j \phi_a)\,,\\
 D_iD_i \phi_a&= \beta (| \phi_a|^2 - 1)\phi_a-A_\alpha A^\alpha\phi_a-
\omega^\alpha(\omega_\alpha-2A_\alpha)\phi_2\delta_{a}^2\, .
\end{align}
\end{subequations}
For configurations satisfying the above Gauss law-type equations (\ref{Gauss-law}a),
the energy per unit length of the string is given by ${\mathcal E}=\eta^2E$, where the
''dimensionless'' energy per unit length, $E$, used throughout this paper can be expressed as:
\be\label{energy-red2}
E=\!\int d^2x\, T^0_{\phantom0 0}=
\!\int d^2x \Bigl\{\omega_\alpha(\omega_\alpha-A_\alpha)|\phi_2|^2
+ \frac{1}{2}B^2+
|D_i \phi_a|^2+\frac{\beta}{2}\left(|\phi_a|^2 - 1\right)^2\Bigr\}\,,
\end{equation}
where $B$ stands for the $z$-component of the magnetic field
(orthogonal to the $(x_1,x_2)$ plane);
$B=B_3=F_{12}=\epsilon_{ij}\pa_iA_j$, and summation over the repeated
$\alpha=0,3$-indices
(not raised with the residual Minkowskian metric) is understood.

The Noether current associated to the residual U(1) symmetry
of the ansatz \eqref{ansatz1}
actually coincides with the $3$-rd isotopic component, $j_\mu^3$,
of the global SU(2) current in Eq.\ \eqref{cur}.
Introducing ${^{(a)}\!J}_\mu\equiv -i(\phi_a\overline{D_\mu\phi}_a-\overline{\phi}_a D_\mu\phi_a)$,
where now no summation is implied over $a=1,2$,
one has $j_\mu^3={^{(1)}\!J}_\mu-{^{(2)}\!J}_\mu$.
On the other hand, the local U(1) current in Eq.\ \eqref{cur} can be written as
$j_\mu^0={^{(1)}\!J}_\mu+{^{(2)}\!J}_\mu$.
Since, by virtue of the field equations \eqref{4d-eqs},
for fields decaying sufficiently fast at infinity the integral of $j_\mu^0$ over the two-plane
vanishes, it follows that the integral of the current $j_\mu^3$ can be expressed as
\be \label{j33}
 \int d^2x\,j_\alpha^3=
2\int d^2x\, {^{(1)}\!J}_\alpha=-2\int d^2x\; {^{(2)}\!J}_\alpha:= 4{\mathcal I}_\alpha,
 \end{equation}
where the ``string worldsheet current'', ${\mathcal I}_\alpha$, is
\be\label{charges}
{\cal I}_\alpha=\int d^2x\, A_\alpha{\bar \phi}_1\phi_1=
\int d^2x\, (\omega_\alpha-A_\alpha){\bar \phi}_2\phi_2\,,
\end{equation}
whose components are the conserved Noether charge per unit length of the string,
${\cal I}_0$, and the total current flowing along the string, ${\cal I}_3$.

Due to the translation symmetry of the Ansatz (\ref{ansatz1}) there is a
conserved momentum, $P$, which is expressed (per unit length) as
\be\label{momentum}
P=\int d^2x\, T^0_{\phantom0 z}  =2\omega_0{\cal I}_3\,,
\end{equation}
and for configurations which possess an additional rotational symmetry in
the plane (which is the case for
all solutions presented in this paper)
one also has a conserved angular momentum,
$J$ (per unit length), given by:
\be\label{angular-momentum}
J=\int d^2x\, T^0_{\phantom0 \varphi}\,.
\end{equation}
Note that just as for the energy per unit length, ${\mathcal E}=\eta^2E$,
to obtain the dimensionful momentum and angular momentum one should multiply $P$ and $J$ by $\eta^2$.
\subsection{Energy bounds and \bogo\ equations}

Let us recall briefly the method, established by \bogo\ \cite{bog},
to find a lower bound for the energy functional \eqref{energy-red2}.
We first introduce the total magnetic flux :
\be\label{flux}
\int d^2x B=\oint A_j dx^j=2\pi n\,,
\end{equation}
with $n$ being an integer winding number.
The flux \eqref{flux} is quantized and topologically conserved,
due to the non-trivial topology of the gauge orbits.
Thus configurations with different values of $n$ are separated by
infinite energy barriers
even though the potential energy cost can be arbitrarily small, since the Higgs field
is topologically trivial \cite{vac-ach}.
In what follows, we shall refer to $n$ as the winding number or number
of magnetic flux quanta. Without any loss of generality we can assume that
$n>0$.
Next, using the identity
\be\label{bogo-id}
\frac{1}{2}|D_i\Phi+ i\epsilon_{ij}D_j\Phi|^2=|D_i\Phi|^2+\epsilon_{ij}\pa_i(\Phi^\dagger D_j\Phi)-B|\Phi|^2\,
\end{equation}
and the expression \eqref{charges} for
${\cal I}_\alpha$,
the energy functional \eqref{energy-red2} becomes
\be\label{bogo-energy}
E=2\pi n+|\omega_\alpha{\cal I}_\alpha|+
\!\int d^2x \Bigl\{
\half|D_i\Phi + i\epsilon_{ij}D_j\Phi|^2 + \half(B+|\Phi|^2-1)^2
+\half(\beta-1)(|\Phi|^2-1)^2
\Bigr\}\,.
\end{equation}
It can now be seen that for the special value $\beta=1$, the energy is
minimized to $2\pi n+|\omega_\alpha{\cal I}_\alpha|$
when the following first order equations, the \bogo\ equations, are satisfied:
\begin{eqnarray}
D_i\Phi + i\epsilon_{ij}D_j\Phi& = & 0, \nonumber \\
B + |\Phi|^2 -1 & = & 0.
\label{bogo-eqs}
\end{eqnarray}
If in a fixed sector of the phase space labeled by the winding number $n$,
the value of $|\omega_\alpha{\cal I}_\alpha|$ can be maintained fixed when varying
the action \eqref{action-red}, then 
solutions of the \bogo\ equations \eqref{bogo-eqs} will solve the variational
Euler-Lagrange
equations \eqref{Gauss-law}.
In the static and untwisted case, when $\omega_\alpha=0$,
this is automatic, but for $\omega_\alpha\ne0$ this is not true in general, except for
solutions with $\omega_\alpha\omega^\alpha=0$ discussed in the next Section.

The known static and untwisted ($\omega_\alpha=0$) semilocal solutions are either
the embedded ANO vortices, with $A_i=A_{{\scriptscriptstyle(\rm ANO)}i}$ and
$\Phi=\phi_{\scriptscriptstyle\rm (ANO)}\Phi_0$, where
 $\Phi_0$ is a constant SU(2) doublet
of unit norm, $\Phi_0^\dagger\Phi_0=1$, or the so-called ``skyrmion'' \cite{hin,Gibbons}
solutions. The latter fulfill
the \bogo\ equations \eqref{bogo-eqs}, and for any winding number, $n$, there exists a
$4n$-parameter family of such solutions describing multi-vortices.
$2n$ of these parameters can be identified with the positions of the vortices ($n$ points in the plane),
and every vortex located at a point $p$ is labeled by a complex parameter, $w_p$, whose
modulus determines the width of the flux tube at point $p$, and whose argument
corresponds to the orientation of the vortex in the ``internal'' space.

\section{Lorentz classes}
\begin{figure}[ht]
    \psfrag{k}{$\omega_3$}
    \psfrag{omega}{$\omega_0$}
    \psfrag{s>0}{$\omega^2>0$}
    \psfrag{ss>0}{$\omega^2>0$}
    \psfrag{s<0}{$\omega^2<0$}
    \psfrag{ss<0}{$\omega^2<0$}
    \psfrag{XPEN}{$\omega^2=0$,~~spinning skyrmions}
    \psfrag{X}{ANO vortices,}
     \psfrag{XX}{skyrmions}
     \psfrag{S}{superconducting}
     \psfrag{SS}{vortices, ${\cal I}>0$}
    \psfrag{S1}{superconducting}
     \psfrag{SS1}{vortices, ${\cal I}<0$}
   \psfrag{h}{rest frame solutions}
     \psfrag{hh}{}
    \resizebox{12cm}{9cm}{\includegraphics{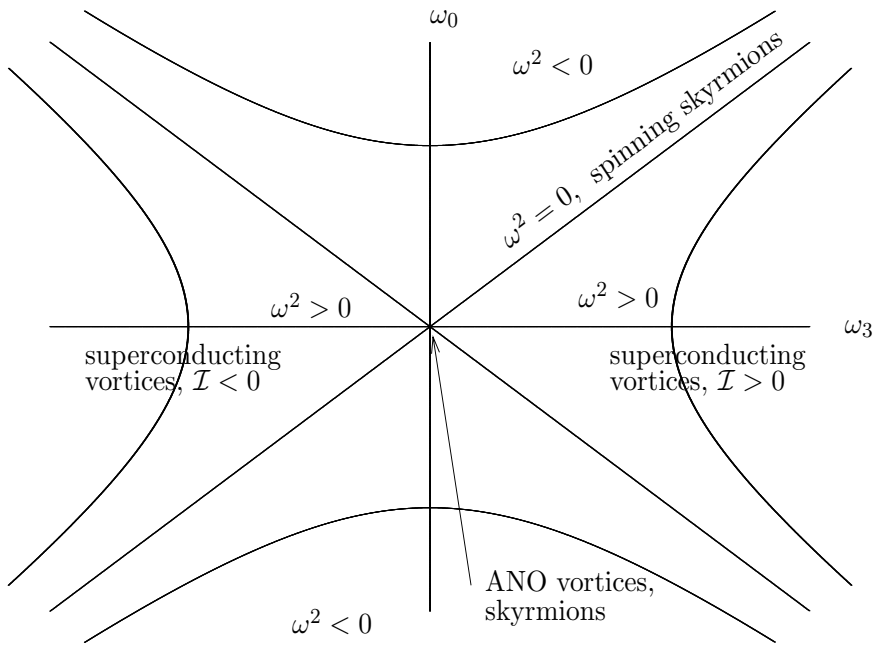}}%
    \caption{Phase space of the stationary and translationally symmetric strings.
Solutions corresponding to a Lorentz orbit with
constant $\omega^2=\omega_3^2-\omega_0^2$
are related to each other by the boost symmetry.
The new superconducting strings correspond to the $\omega^2>0$ regions.
}
\label{fig1}
\end{figure}

\setcounter{equation}{0}
We notice at this point that our coordinate system is not yet completely fixed,
since one can perform  boosts along
the $z$-axis:
\be\label{boost}
x^\mu=(t,z,x_1,x_2)\to \tilde{x}^{\prime\mu}
=(\tilde{t},\tilde{z},x_1,x_2),
\end{equation}
where
\be\label{boost1}
t=\tilde{t}\cosh(\gamma)-\tilde{z}\sinh(\gamma),~~~~
z=\tilde{z}\cosh(\gamma)-\tilde{t}\sinh(\gamma),
\end{equation}
with $\gamma$ being the parameter of the boost. The fields then transform
as
\be\label{boost-fields}
A_\mu(x)dx^\mu=\tilde{A}_\mu(\tilde{x})d\tilde{x}^\mu,~~~~
\Phi(x)=\tilde{\Phi}(\tilde{x})\,,
\end{equation}
and the form of our Ansatz (\ref{ansatz1}) will be preserved upon replacing
$t,z$ by $\tilde{t},\tilde{z}$ and $(\omega_0$, $\omega_3)$ by
\be\label{boost-omega}
\tilde{\omega_0}=\omega_0\cosh(\gamma)-\omega_3\sinh(\gamma)\,,\quad
\tilde{\omega_3}=\omega_3\cosh(\gamma)-\omega_0\sinh(\gamma)\,.
\end{equation}
We thus see that Lorentz boosts in the $(t,z)$ plane preserve the structure of the
ansatz (\ref{ansatz1}), and that the
$\omega_\alpha$ transform as components of a Lorentz (co)vector.
This means that all pairs of values of $\omega_\alpha$
belonging to the same orbit of the Lorentz group (see Fig. \ref{fig1})
are equivalent. As a result,
instead of two independent parameters $(\omega_0$, $\omega_3)$
we have in fact only one, which can be chosen to be
the Lorentz-invariant combination
\be
\omega^2=-\omega^\alpha\omega_\alpha=\omega_3^2-\omega_0^2\,.
\end{equation}
Therefore the space of solutions
decomposes into three classes that can be labeled by
the possible Lorentz types of the length of $\omega^2$,
as proposed by Carter \cite{Carter}:
\be
\omega^2\begin{cases}
 =0&\text{null or chiral case}\\
 <0&\text{time-like or electric case}\\
 >0&\text{space-like or magnetic case}
\end{cases}
\end{equation}

These three classes also correspond to the Lorentz types of the string worldsheet
current ${\cal I}_\alpha$ in \eqref{charges}.
We now discuss some relevant properties of Eqs.\ (\ref{Gauss-law}) corresponding
to the three Lorentz classes.
\begin{enumerate}
 \item The chiral class, $\omega_\alpha\omega^\alpha=0$.
All known semilocal solutions obtained up to now belong to this class.
From Eqs.\ (\ref{Gauss-law}a) it follows that
$A_{\scr(\omega)}:=\omega^\alpha A_\alpha$ satisfies
\be \label{A0}
\triangle A_{\scr(\omega)}-2A_{\scr(\omega)}\phi_a\overline{\phi}_a=0.
\end{equation}
 Now the standard positivity argument for Eq.\ (\ref{A0}),
\be\label{A01}
 \int d^2x  A_{\scr(\omega)}\triangle A_{\scr(\omega)}=-\int d^2x (\pa_iA_{\scr(\omega)})^2=
 2\int d^2x(A_{\scr(\omega)})^2\phi_a\overline{\phi}_a\,,
\end{equation}
implies that the only regular and bounded solution of Eq.\ (\ref{A0}) in the 2-plane is the
 trivial one, $A_{\scr(\omega)}\equiv0$, i.e.\ $\omega_0A_0=\omega_3A_3$.
Therefore the two Gauss-law type equations (\ref{Gauss-law}a) reduce to a single one.
This also implies that the worldsheet current is isotropic,
${\cal I}_\alpha {\cal I}^\alpha=0$.
Furthermore, since $A^\alpha A_\alpha=0$,
Eqs.\ (\ref{Gauss-law}b,c) reduce to the equations of motion of the {\sl untwisted} and
{\sl static} semilocal vortices, and they {\sl decouple} from the remaining Gauss law.
Therefore, to any static, untwisted vortex solution $(A_i,\phi_a)$
of Eqs.\ (\ref{Gauss-law}b,c)
there corresponds a twisted, stationary
chiral vortex obtained by solving in addition
the  Gauss-law equation (\ref{Gauss-law}a). It is important to note that
the solutions $A_\alpha$ of the Gauss-law equation (\ref{Gauss-law}a)
do not always fall off sufficiently fast at infinity. As a result, not all
of the stationary chiral solutions will have finite energy.

As already mentioned, the known static and untwisted semilocal solutions
of Eqs.\ (\ref{Gauss-law}b,c) are either
the embedded ANO vortices, or the ``skyrmions''.
Bounded solutions of the related Gauss-law equation (\ref{Gauss-law}a)
exist only for the ''skyrmions'', but
their energy for winding number $n=1$ is  nevertheless divergent
due to the slow falloff of the fields $A_\alpha$ (lack of localization).
This is an alternative way to rederive the main result of Ref. \cite{Abraham},
where stationary and charged vortices satisfying the \bogo\ equations were found,
to which we shall refer to as ``spinning skyrmions''.

\item  The space-like (or magnetic) class, $\omega_\alpha\omega^\alpha<0$.
In fact all new solutions to be presented in this paper fall into this class.
Now contracting Eqs.\ (\ref{Gauss-law}a) with $\tilde\omega_\alpha$ where
$\tilde\omega^\alpha \omega_\alpha=0$ one finds that
$A_{\scr(\tilde\omega)}:=\tilde\omega^\alpha A_\alpha=\omega_3A_0-\omega_0A_3$ satisfies
\be \label{A0tilde}
\triangle A_{\scr(\tilde\omega)}-2A_{\scr(\tilde\omega)}\phi_a\overline{\phi}_a=0\,.
\end{equation}
The same positivity argument as above implies again $A_{\scr(\tilde\omega)}\equiv0$,
i.e.\ the electric potential, $A_0$, is completely determined by
the component of the magnetic one, $A_3$.
Exploiting the boost-symmetry of the Ansatz (\ref{ansatz1}),
one can always achieve in this case by a choice of the reference frame
that $\omega_0=0$ and $A_0=0$. We note that the worldsheet current is spacelike,
${\cal I}_\alpha {\cal I}^\alpha<0$.
\item  The timelike (or electric) class, $\omega_\alpha\omega^\alpha>0$.
In this case there is such a  reference frame
where $\omega_3=0$ and $A_3=0$. The worldsheet current is timelike,
${\cal I}_\alpha {\cal I}^\alpha>0$.
It is unlikely that finite energy solutions exist in this class, since from
Eqs.\ (\ref{Gauss-law}c)
we obtain the asymptotic $(\rho^2=x_1^2+x_2^2\to\infty)$ equation
$\triangle f_2=-\omega_\alpha\omega^\alpha f_2$
(assuming that $f_1\to1$, $A_\mu\to0$), which implies that the function $|f_2|$ tends
too slowly to zero to make the energy \eqref{energy-red2} convergent.
\end{enumerate}

\section{The Cylindrically symmetric equations}
 \setcounter{equation}{0}

\subsection{Field equations}

In this section we shall examine solutions with cylindrical symmetry, for which we may choose
a gauge such that we are lead to the following Ansatz in polar coordinates:
 \begin{align}\label{ansatz2}
&A_0=\omega_0a_0(\rho)\,,A_\rho=0\,, A_\varphi=na(\rho)\,,
A_3=\omega_3 a_3(\rho)\,,\notag\\
&\phi_1 = f_1(\rho)e^{in\varphi}, \quad
\phi_2 = f_2(\rho)e^{im\varphi}e^{i(\omega_0 t+\omega_3 z)}\,,
\end{align}
where $m=0,\ldots n-1$. Furthermore, as shown in the previous section,
the electric potential is either given by $A_0=A_3$ in the chiral case ($|\omega_0|=|\omega_3|$),
or by $A_0=\omega_0A_3/\omega_3$ in the magnetic case, i.e.\ in both cases one can take
$a_0(\rho)=a_3(\rho)$.
The energy functional assumes the form:
\be\label{energy}
\begin{split}
E=
&2\pi\int_0^\infty \rho d\rho\,T^0_ {\phantom0 0}=
2\pi\int_0^\infty \rho d\rho\biggl\{
\frac{1}{2}\Bigl(n^2\frac{{a'}^2}{\rho^2}+|\omega|^2{a_3'}^2\Bigr)
+n^2\frac{(1-a)^2}{\rho^2}f_1^2 \\
&+\frac{(m-na)^2}{\rho^2}f_2^2+
{f_1'}^2+{f_2'}^2+|\omega|^2\bigl[ f_1^2a_3^2+f_2^2(1-a_3)^2\bigr]
+\frac{\beta}{2}\left(1-|f|^2\right)^2 \biggr\}\,,
\end{split}
\end{equation}
where $|\omega|^2\equiv\omega_0^2+\omega_3^2$, $|f|^2\equiv f_1^2+f_2^2$.
A \bogo-type rearrangement of the energy \eqref{energy} yields the following:
 \be\label{cylindr-bogobound}
\begin{split}
&E= 2\pi n + |\omega^2|{Q} +
\pi(\beta-1)\int\limits_0^\infty \rho d\rho\,\bigl(1-|f|^2)^2
+ \\
&2\pi\int\limits_0^\infty\rho d\rho\biggl\{
\frac{1}{2}\left(\frac{na'}{\rho}+(|f|^2-1)\right)^2+\left({f_1}'+
\frac{n(a-1)}{\rho}f_1\right)^2
+\left({f_2}'+\frac{(na-m)}{\rho}f_2\right)^2 \biggr\}\,,
\end{split}
\end{equation}
where the quantity $Q$,
 \be\label{rescaled-current}
{Q}=2\pi\int\limits_0^\infty\rho d\rho\,(1-a_3)f_2^2 =
2\pi\int\limits_0^\infty\rho d\rho\, a_3f_1^2\,,
\end{equation}
determines the vortex worldsheet current,
\be\label{III}
\mathcal{I}_\alpha=\omega_\alpha{Q},
\end{equation}
and in terms of which the momentum \eqref{momentum} and the angular momentum
\eqref{angular-momentum}
can be expressed as
\begin{eqnarray}                \label{PJ}
P&=&2\omega_0\omega_3 Q\,, \nonumber  \\
J&=&-2\omega_0\nu Q\,,\quad {\rm where\ } \nu:= n-m=1,\ldots,n\,.
\end{eqnarray}
The parameter $Q$ is Lorentz-invariant,
since it enters the invariant relation (\ref{III}) between
Lorentz-vectors. On the other hand, the quantities
$\mathcal{I}_\alpha,P,J$ are not Lorentz-invariant. In particular,
in the spacelike case, if $\omega^\alpha\omega_\alpha<0$, there
exists a rest frame where $\omega_0=0$, therefore
in this frame $\mathcal{I}_0=P=J=0$.  Using Eq.\eqref{PJ}, we can
define in any nonstatic frame a ``helicity'' or ``polarization''
given by the ratio
\be
\label{hel} \frac{J}{P}=-\frac{\nu}{\omega_3}\,,
\end{equation}
which can be extended as a limit to the static reference frame as well.
From now on we shall refer to $\nu$ as helicity number.
In the rest frame, where $\omega_3=\omega$, we define the ``rest frame current'' as
\be\label{restframecurr}
\mathcal{I}:=\omega Q\,.
\end{equation}
The value of the energy in the rest frame is given by Eq.\ \eqref{cylindr-bogobound} with
$|\omega|^2=\omega^2$.

Eq. \eqref{cylindr-bogobound} implies the following inequality for $\beta\geq 1$:
\be\label{bound}
E\geq 2\pi n+(\omega_0^2+\omega_3^2){Q}\,.
\end{equation}
It is worth noting that the existence of this bound does not
automatically guarantee the existence of stable, localized solutions
in the model \cite{Preskill}.
The bound (\ref{bound}) can be saturated
for $\beta=1$, if all the perfect squares in the
second line in (\ref{energy}) vanish, leading to the \bogo\ equations:
\begin{subequations}\label{cyl-bogoeqs}
\begin{align}
{f_1}'+\frac{n(a-1)}{\rho}f_1& = 0 \,,\\
{f_2}'+\frac{(na-m)}{\rho}f_2& = 0 \,,\\
\frac{na'}{\rho}+(f_1^2+f_2^2-1)& = 0\,.
\end{align}
\end{subequations}
The reduced action, ${\cal S}$,
can be obtained by replacing in the expression for the energy $E$ in
 \eqref{energy} $|\omega|^2=\omega_3^2+\omega_0^2$ by
$\omega^2=\omega_3^2-\omega_0^2$. Varying ${\cal S}$ gives the
cylindrically symmetric field equations:
\begin{subequations}\label{cyl-eqs}
\begin{align}
\frac{1}{\rho}(\rho a_3')'& =2a_3|f|^2-2f_2^2 \,,\\
 \rho{\biggl(\frac{a'}{\rho}\biggr)}'& = 2f_1^2(a-1)+2f_2^2(a-\frac{m}{n}) \,,\\
 \frac{1}{\rho}(\rho f_1')'& = f_1\left[n^2\frac{(1-a)^2}{\rho^2}+\omega^2a_3^2-
\beta(1-|f|^2)\right],\\
 \frac{1}{\rho}(\rho f_2')'& = f_2\left[\frac{(m-na)^2}{\rho^2}+\omega^2(1-a_3)^2-
\beta(1-|f|^2)\right]\,.
\end{align}
\end{subequations}
where $'=d/d\rho$. These equations do not depend separately
on $\omega_0$
and $\omega_3$, but only on the Lorentz-invariant
combination $\omega^2=\omega_3^2-\omega_0^2$, and therefore any solution
determines actually a whole class, which is its Lorentz orbit corresponding to
boosts \eqref{boost-fields}.
Since we are interested
in finite energy solutions of Eqs.\ \eqref{cyl-eqs},
in the following we shall assume that
$\omega^2\geq0$ (space-like or null classes in the terminology of the previous Section).

We note that the change of variable $a(\rho)\to n(1-a(\rho))$
achieves that the field equations do not depend separately on $n$
and $m$, but only on the ``helicity number'' $\nu=n-m$.
However, due to the boundary conditions, solutions of the field equations do not
depend only on $\nu$, but separately on $n,m$. For this reason we
prefer to use the parameterization of the equations explicitly
containing both $n$ and $m$.

One can directly check that any solution of the first order
\bogo\ equations \eqref{cyl-bogoeqs} also solves the field equations \eqref{cyl-eqs} if $\beta=1$.
In the next Section we shall briefly review some relevant
solutions of the \bogo\ equations \eqref{cyl-bogoeqs}.

\subsection{Local behaviour}

Eqs.\ (\ref{cyl-eqs}) admit a $4$-parameter family of local solutions regular
at the origin, $\rho=0$:
\begin{subequations}\label{ori}
\begin{align}
 a& =a^{(2)}\rho^2+O(\rho^{2m+2})\,,&\quad a_3& = a_3^{(0)}+O(\rho^{2m+2})\,,&\\
f_1 &=f_1^{(n)}\rho^n+O(\rho^{n+2})\,,&\quad f_2& = f_2^{(m)}\rho^m+O(\rho^{m+2})\,,&
 \end{align}
 \end{subequations}
where $a^{(2)}$, $a_3^{(0)}$, $f_1^{(n)}$, $f_2^{(m)}$
are free parameters.
Depending on the relative magnitudes of the values of the parameters
$\beta$ and $\omega\equiv\sqrt{\omega^2}$ with respect to $\sqrt{2}$,
the various possible asymptotic
behaviours of the fields as $\rho\to\infty$ for $\omega>0$ can be
inferred from the following:
\begin{subequations}\label{inf}
\begin{align}
a& = 1+A\sqrt{\rho}\,e^{-\sqrt{2}\rho}-
\frac{D^2(1-m/n)}{(1-2\omega^2)}\frac{e^{-2\omega\rho}}{\rho}+\ldots\,,\\
a_3& = B\frac{e^{-\sqrt{2}\rho}}{\sqrt{\rho}}+
\frac{D^2}{(1-2\omega^2)}\frac{e^{-2\omega\rho}}{\rho}+\ldots\,,
\\
f_1& = 1+C\frac{e^{-\sqrt{2\beta}\rho}}{\sqrt{\rho}}
-\frac{\beta D^2}{2(\beta-2\omega^2)}\frac{e^{-2\omega\rho}}{\rho}
+\frac{n^2A^2+\omega^2B^2}{2(4-\beta)}\frac{e^{-2\sqrt{2}\rho}}{\rho}+
\ldots\,, \\
f_2& = D\frac{e^{-\omega\rho}}{\sqrt{\rho}}\, +\ldots~,
\end{align}
 \end{subequations}
where $A,B,C,D$ are free parameters, and the resonant values
($\omega^2=1/2$, etc) have been excluded.
The dominant asymptotic behaviour of $a$, $a_3$, $f_1$ is
given by the smallest exponential in \eqref{inf}.

If $\omega^2=0$ then the Gauss-law-type equation for $a_3$ (\ref{cyl-eqs}a)
(recall that $a_0=a_3$)
decouples
from the field equations (\ref{cyl-eqs}b-d).  The latter then become identical to the
much studied ``minimal'' (i.e. {\sl static} and {\sl untwisted})
semilocal vortex equations.
Let us recall some known facts about this ``minimal''
semilocal system (\ref{cyl-eqs}b-d).
If $f_2=0$, then
Eqs.\ (\ref{cyl-eqs}b-c) reduce to the vortex equations in the N$=1$ Abelian-Higgs model,
in which case solutions are known to exist for any value of $\beta$.
The behaviour of these ANO vortices near the singular points is immediately obtained from
Eqs.\ \eqref{ori} and \eqref{inf} (with $B=D=0$). In addition, 
Hindmarsh \cite{hin} and Gibbons et al. \cite{Gibbons}
have shown the existence of $n$ one-parameter families of solutions of
Eqs.\ (\ref{cyl-eqs}b-d) also with $f_2\neq 0$, but only for the special value $\beta=1$. In fact,
these solutions actually fulfill 
the first order \bogo\ equations \eqref{cyl-bogoeqs},
and they are sometimes referred to as ``skyrmions''.
The asymptotic behaviour for large $\rho$ of these solutions is radically
different from that of the
ANO vortices, whose fields approach their vacuum values exponentially,
whereas the ``skyrmions''
exhibit an asymptotic power law behaviour. In fact for $\omega=0$ the asymptotic
behaviour of the fields in Eqs.\ \eqref{cyl-eqs} is given as:
\bea\label{inf-skyrm}
a&=&1+
\left(-\frac{(1-m/n)}{x^{2\nu}}+\ldots\right)+A\sqrt{\rho}\,e^{-\sqrt{2}\rho}+\ldots\,,
\nonumber \\
a_3&=&
\left(\frac{1}{x^{2\nu}}+\ldots\right)+B\frac{e^{-\sqrt{2}\rho}}{\sqrt{\rho}}+\ldots\,,
\nonumber
\\
f_1&=&1+
\left(-\frac{1}{2}\frac{1}{x^{2\nu}}+\ldots\right)
+C\frac{e^{-\sqrt{2\beta}\rho}}{\sqrt{\rho}}+\ldots\,,\nonumber \\
f_2&=&\frac{1}{x^{\nu}} +\ldots~,
\eea
where $x\equiv\rho/w$ with $w$, the width of the magnetic flux tube,
being an arbitrary parameter.
The magnetic field, $B$, falls off like
$w^{2\nu}/\rho^{2\nu+2}$.
This is quite contrary to what
one could expect, namely that the flux be confined within a region of
size $\approx1/\sqrt{2}$
determined by the length scale of the massive vector boson.
It should be stressed at this point that although
the asymptotic expansion \eqref{inf-skyrm}
is valid for any value of $\beta$,
{\sl global solutions} of \eqref{cyl-eqs}
for $\omega=0$ satisfying the vortex boundary
conditions \eqref{ori},\eqref{inf-skyrm} are known
to exist only for $\beta=1$.

In this paper we shall give convincing numerical evidence
that by relaxing the condition $\omega=0$,
$n$ distinct one-parameter families of
(twisted) solutions of Eqs.\ \eqref{cyl-eqs}
exist not only for $\beta=1$, but also for any value of $\beta>1$.

\section{Untwisted and chiral Semilocal vortices}
 \setcounter{equation}{0}

Before considering new solutions, it is instructive to reproduce numerically
some of the known solutions of the field equations (\ref{cyl-eqs})
and to briefly review some of their properties.
The numerical method we have used is a suitable
version of the ``multi-shooting'' procedure
successfully applied to produce high precision data for
the 't Hooft-Polyakov monopoles \cite{FOR}, to which paper we refer for further details.

The simplest solutions are the  ANO vortices corresponding to
$a_0=0$, $a_3=0$, $f_2=0$.
We exhibit a sample ANO vortex profile function, $a=a_{\rm ANO}$, $f_1=f_{\rm ANO}$, in
Fig.\ 2, and also tabulate
some numerical values of the relevant parameters of the ANO solutions in Table I.
The energy of the ANO vortices increases monotonically with $\beta$
and diverges in the London limit
$\beta\to\infty$.
Solutions with $\beta=1$ are distinguished by the fact that they
also satisfy the first order Bogomol'nyi equations \eqref{cyl-bogoeqs}.

\begin{figure}[h]
\begin{center}
\hbox to\linewidth{\hss%
  \psfrag{x}{$\ln(1+\rho)$}
  \psfrag{X}{$\beta=2$}
  \psfrag{f}{$f_{\rm ANO}$}
  \psfrag{a}{$a_{\rm ANO}$}
  \resizebox{8cm}{6.5cm}{\includegraphics{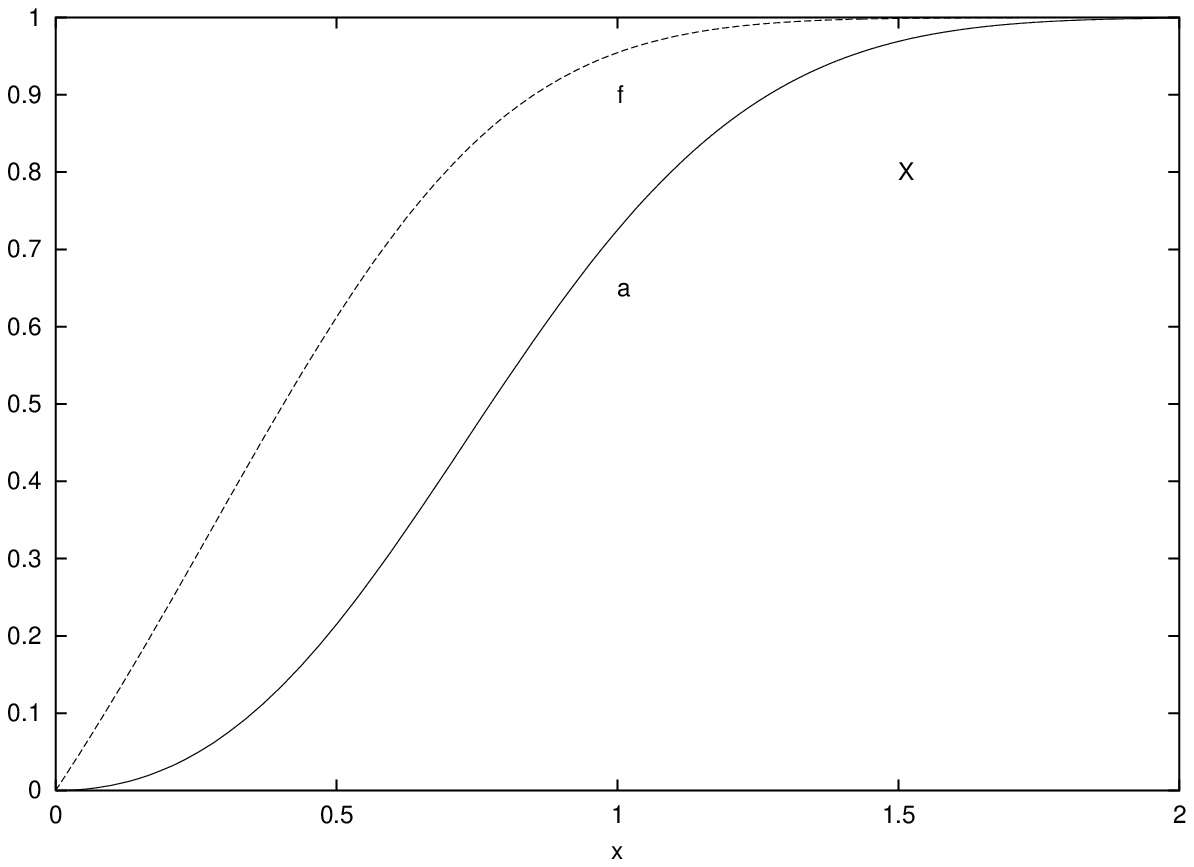}}%
\hspace{5 mm}
  \psfrag{x}{$\ln(1+\rho)$}
  \psfrag{f1}{$f_1$}
  \psfrag{f2}{$f_2$}
  \psfrag{a}{$a$}
  \resizebox{8cm}{6.5cm}{\includegraphics{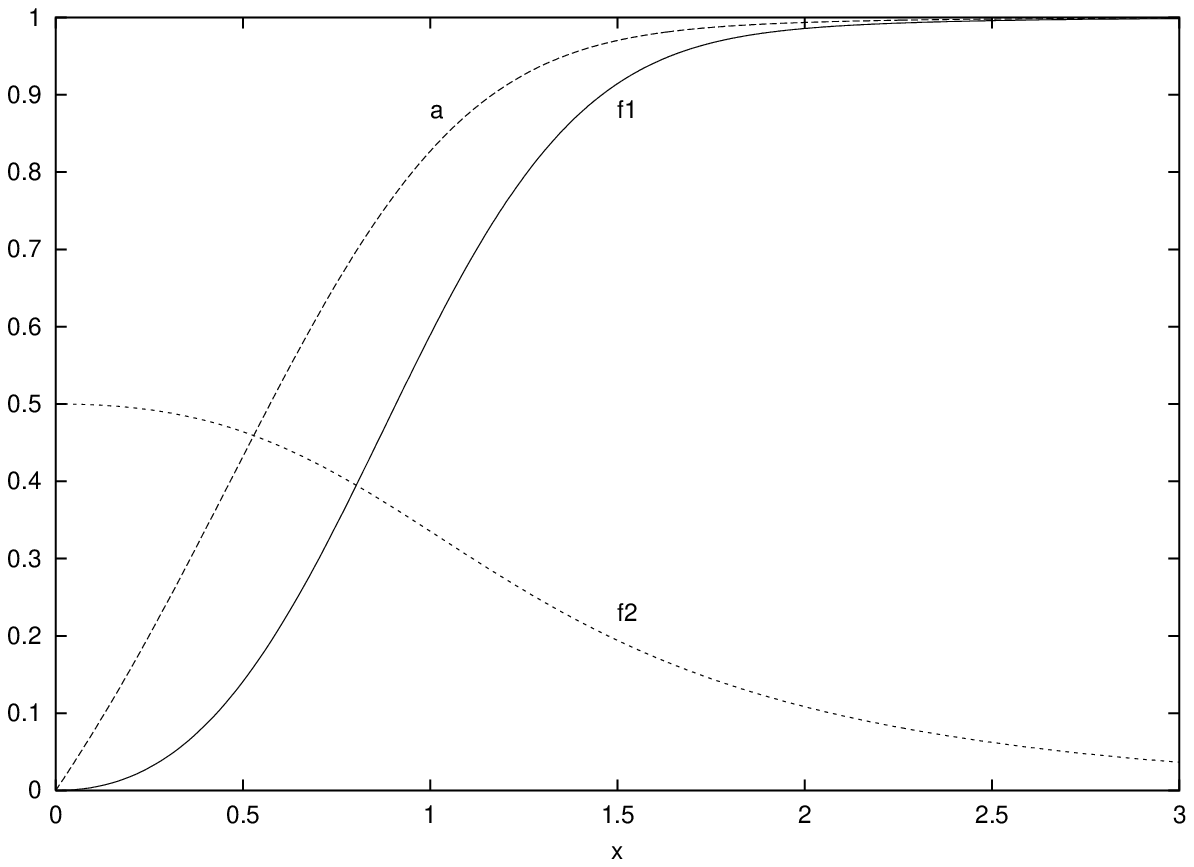}}%
\hss}
\caption{The profile functions for the ANO vortex with $\beta=2$, $n=1$ (left)
and for one of the ``skyrmions'' with $n=1$, $m=0$, $(\beta=1)$ (right).
}
\end{center}
\label{figANO}
\end{figure}
%

Next, as already discussed, for the special value $\beta=1$ there are $n$ one-parameter
families of solutions with $f_2\neq 0$
generalizing the corresponding ANO vortex. These solutions,
dubbed ``skyrmions'',
satisfy the \bogo\ equations \eqref{cyl-bogoeqs}, therefore they all have the
same energy. As noted in \cite{hin},
 the three \bogo\ equations imply that
\be\label{f_2}
f_2=\frac{1}{x^{\nu}}f_1\,,
\end{equation}
with $x={\rho}/{w}$,
so that \eqref{cyl-bogoeqs} reduce to two equations for $f_1$ and $a$.
For a given value of the winding number $n$, the helicity number $\nu=n-m$ can assume
$n$ values, $\nu=1,2,\ldots n$, which label
$n$ different solutions of the Bogomol'nyi equations for any given value
of the ``size parameter'' $w$.
The existence of these solutions was established in
\cite{hin,Gibbons}.
We exhibit a typical solution in
Fig. 2.

In the limit $w\to 0$ the ``skyrmion'' solution tends smoothly
to the $\beta=1$ ANO vortex.
Since we shall make a connection
with the large size limit of the ``skyrmions'', $w\to\infty$,
we recall some of the main points
of the analysis of Ref.\ \cite{hin,Gibbons}.
To study this limit, it is convenient to introduce the variable
\be
\chi=\ln(f)+\frac{1}{2}\ln(1+x^{-2\nu}),
\end{equation}
which satisfies the equation
\be\label{chi-eq}
\frac{d^2\chi}{dx^2}+\frac{1}{x}\frac{d\chi}{dx}+w^2\bigl(1-e^{2\chi}\bigr)=
2\nu^2\frac{x^{2\nu-2}}{(x^{2\nu}+1)^2}\, .
\end{equation}
Now one can see that for large values of $w$
the solution of Eq.\ \eqref{chi-eq} is well approximated by
\be\label{chi-approx}
\chi\approx-\frac{\nu^2}{w^2}\frac{x^{2\nu-2}}{(x^{2\nu}+1)^2},
\end{equation}
which tends to $\chi=0$ as $w\to\infty$,
so as the width of the vortex tends to infinity the ``skyrmion'' solution approaches
the following configuration more and more:
\be\label{skyrm-limit}
f_1=\frac{x^{\nu}}{\sqrt{1+x^{2\nu}}}\,,\quad
f_2=\frac{1}{\sqrt{1+x^{2\nu}}}\,,\quad
n(1-a)=\frac{\nu}{{1+x^{2\nu}}}\,.
\end{equation}
For the limiting configuration \eqref{skyrm-limit} one has
$f_1^2+f_2^2=1$, which means
 that the amplitude of the Higgs field is fixed,
$\Phi^\dagger\Phi=1$. Taking into account the U(1) ``phase symmetry''
one realizes that the Higgs field $\Phi$ takes its values
in the target space of a $\mathbf{CP}^{\scr 1}$ nonlinear $\sigma$-model.
For this reason the ``skyrmions'' in the infinite width limit approximate
$\mathbf{CP}^{\scr 1}$-lumps \cite{hin}.
It is worth pointing out here that the non-trivial part of the limiting configuration
\eqref{skyrm-limit} depends only on $\nu=n-m$. For a fixed value of $\nu$
there exist an infinite number of different ``skyrmions'' with different winding numbers
$n=\nu,\nu+1,\ldots$ whose energies are
different. Although solutions with $n>\nu$ do not have the same boundary
conditions at the axis as (\ref{skyrm-limit}),
away from the origin
all of them approach for $w\to\infty$ the same configuration (\ref{skyrm-limit}).

Finally one may wonder about the possible existence of such
``skyrmions'' with $\omega=0$ also for $\beta\ne1$.
Just by counting the number of free parameters: there are four at $\rho=0$ in \eqref{ori}
and also four at $\rho=\infty$ in \eqref{inf-skyrm},
one sees that one cannot hope to find a one parameter family of solutions similar to
those in the
$\beta=1$ case.
Although
this parameter counting argument leaves open a possibility of finding some isolated
``skyrmion'' solutions, here we can only confirm the negative results of Ref. \cite{hin},
since our numerical searches yielded no solutions with $\omega\neq 0$ for $\beta\ne1$.

\section{Twisted String Solutions }
\setcounter{equation}{0}
In this Section we
present some numerically constructed solutions of the cylindrically
symmetric field equations \eqref{cyl-eqs} for $\beta>1$.
For a given value of the winding number $n$,
the new solutions comprise $n$ one parameter families
labeled by the twist parameter, $\omega$.
Exploiting the boost symmetry gives finally two-parameter families of twisted vortices
labeled by $\omega$ and $\omega_0$.
We also explore the phase space of these twisted vortices, and we believe that
our mostly numerical results provide strong evidence in favour of their existence.
In order to obtain numerical solutions by our shooting and matching techniques it is
necessary to have a ``reasonable'' starting point in the $8$-dimensional parameter space.
In particular one
has to avoid that the numerical procedure converges to the ANO solution, which is always present.
A very natural idea is to search for new classes of solutions bifurcating with the ANO vortex.
If such bifurcating solutions exist, they will
provide a good starting point in the parameter space.

\subsection{Bifurcation analysis}
The first step in the bifurcation analysis is to linearize
Eqs.\ (\ref{cyl-eqs}) around the ANO solution with
$a=a_{\scriptscriptstyle\rm ANO}$, $f_1=f_{\scriptscriptstyle\rm ANO}$,
$a_3=0$, $f_2=0$. If now $f_2\ll 1$, then to 
first order in the amplitude $f_2$ one has $a_3\equiv0$,
since $f_2=0$ implies $a_3=0$,
and $f_2$ appears quadratically in Eq.\ (\ref{cyl-eqs}a). Therefore $a_3=O(f_2^2)$, and
the linearized field equation (\ref{cyl-eqs}d) for the $f_2$ amplitude
can be written as the following (Schr\"odinger-type) equation:
\be\label{Schr}
 -\frac{1}{\rho}(\rho f_2')' +\left[\frac{(na_{\scriptscriptstyle\rm ANO}-m)^2}{\rho^2}
-\beta(1-f_{\scriptscriptstyle\rm ANO}^2)\right]f_2=-\omega^2f_2\,.
\end{equation}
This equation corresponds
precisely to the eigenvalue problem studied by Hindmarsh in his
linearized stability analysis of the ANO vortex
embedded into the semilocal theory
 \cite{hin}. In agreement
with his results, for any $\beta>1$ we find a single normalizable solution, $f_2$,
with a negative eigenvalue, $-\omega^2\equiv-\omega_{\rm b}^2(\beta,n,m)$.
In Table 1\ we list for
$n=1$, $m=0$ the values of $\omega_{\rm b}(\beta)$ for some values of $\beta$.
Just as in Ref.\ \cite{hin}
we do not find any non-trivial normalizable solutions of Eq.\ (\ref{Schr})
when $\beta<1$.
\begin{table}
\begin{center}
\parbox{12cm}{\caption{Numerical values of the shooting
parameters $f_1^{(1)}$ and $a^{(2)}$,
the energy for the ANO vortex and
the bifurcation parameter $\omega_{\rm b}$ corresponding to the branching off of
 the twisted solutions.
}} \vskip 0.3cm
\begin{tabular}{ccccccc}
$\beta$ & $\omega_{\rm b}(\beta)$ & $f_1^{(1)}$ & $a^{(2)}$ & ${E}/2\pi$\\
\hline
1  & 0        & 0.500000 & 0.853178 & 1.000000\\
2  & 0.329886 & 0.616573 & 1.099352 & 1.156761\\
3  & 0.513051 & 0.696864 & 1.284290 & 1.260901\\
4  & 0.661658 & 0.759706 & 1.438747 & 1.340595\\
5  & 0.790405 & 0.811948 & 1.574136 & 1.405773\\
6  & 0.905687 & 0.856955 & 1.696163 & 1.461215\\
7  & 1.011028 & 0.896661 & 1.808164 & 1.509620\\
8  & 1.108629 & 0.932290 & 1.912281 & 1.552676\\
9  & 1.199978 & 0.964672 & 2.009991 & 1.591515\\
10 & 1.286135 & 0.994398 & 2.102359 & 1.626935\\
\hline
\end{tabular}
\end{center}
\end{table}
In Ref.\ \cite{hin}, where $\omega_3=0$,
the existence of normalizable solutions (bound states)
of Eq.\ (\ref{Schr}) for $\beta>1$ is the sign of the instability of the embedded
ANO vortex, whereas their absence for $\beta<1$ indicates its linearized stability.
The instability of the embedded ANO vortex
can be interpreted as the magnetic flux spreading out to infinity.

\begin{figure}[ht]
\hbox to\linewidth{\hss%
\psfrag{beta}{$\beta$}
  \psfrag{X}{\large$\omega_{\rm b}^2(\beta,n,m)$}
  \psfrag{n2,m=0}{\hspace*{-2cm}$n=2,\,m=0$}
  \psfrag{n1,m=0}{\hspace*{-2cm}$n=1,\,m=0$}
  \psfrag{n2,m=1}{$n=2,\,m=1$}
  \resizebox{8cm}{6.5cm}{\includegraphics{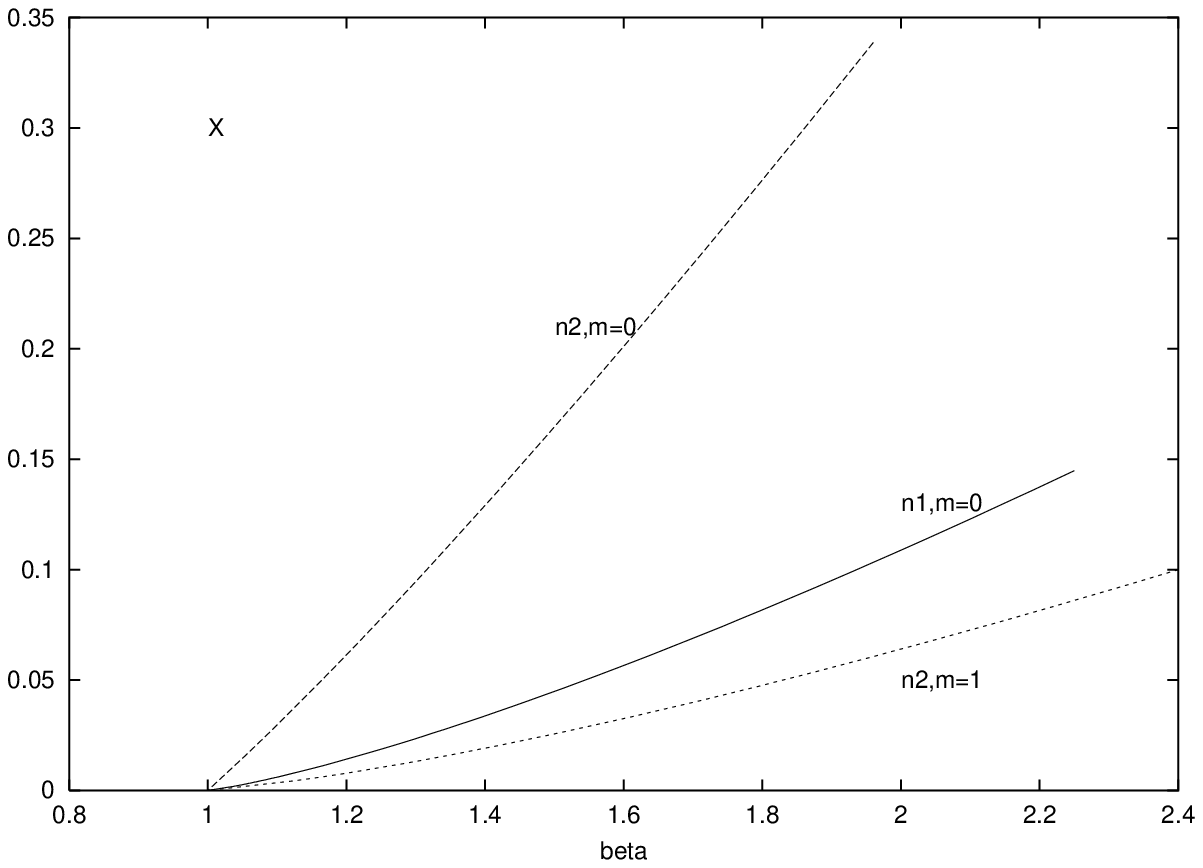}}%
\hspace{5mm}%
  \psfrag{q}{$f_2^{(0)}$}
  \psfrag{sigma}{$\omega$}
  \psfrag{b=2}{$\beta=2$}
  \psfrag{b=4}{$\beta=4$}
  \psfrag{b=9}{$\beta=9$}
    \resizebox{8cm}{6.5cm}{\includegraphics{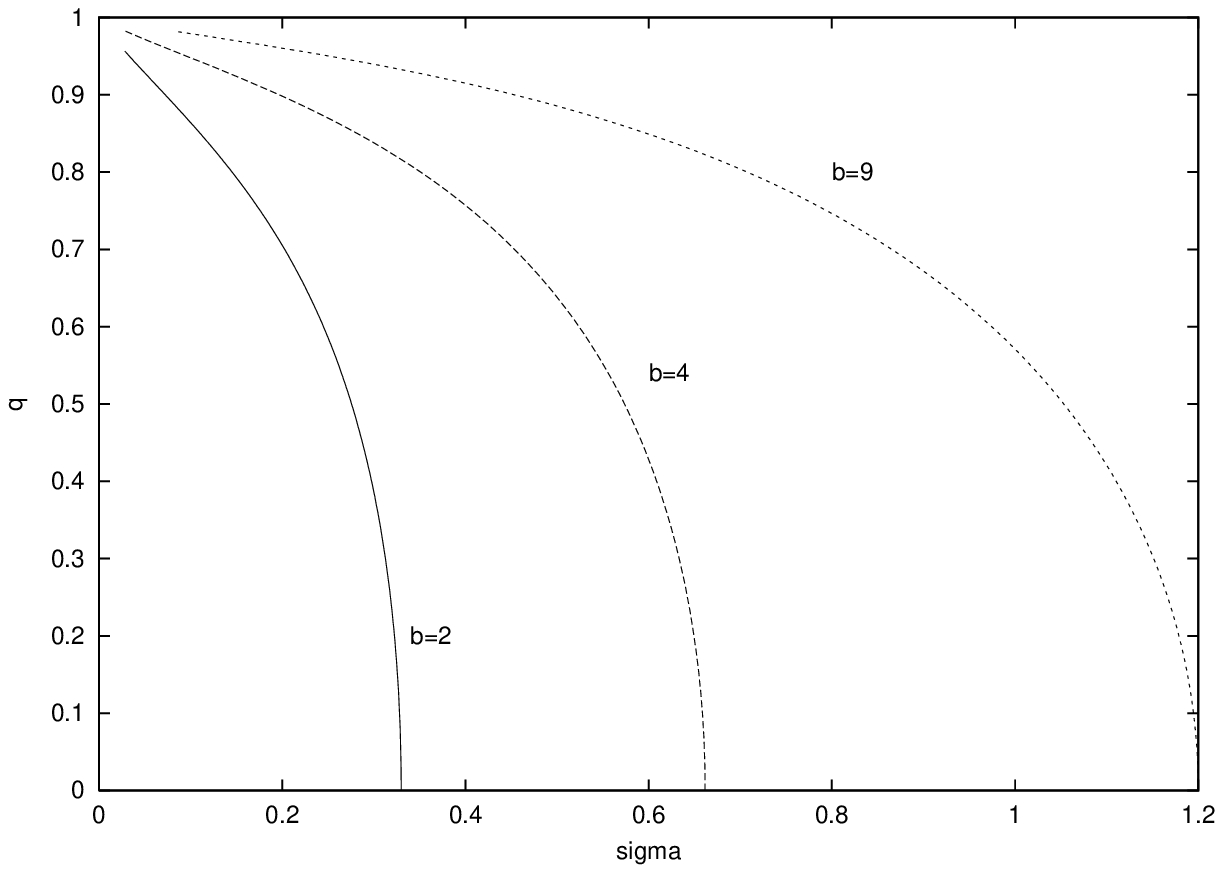}}%
\hss}
\caption{Left: The bifurcation parameter, $\omega_{\rm b}^2(\beta,n,m)$, as a
function of $\beta$.
Right: the value of the condensate $f_2^{(0)}$ as a function of the twist
$\omega$
for twisted solutions with $\beta=2,\,4,\,9$, and $n=1$. }
\label{fig4}
\end{figure}

In our case $\omega_3\neq 0$, and the very same bound states indicate the existence
of new branches of solutions bifurcating with
the corresponding ANO vortex at $\omega=\omega_{\rm b}(\beta,n,m)$.
The bifurcation parameter
$\omega_{\rm b}(\beta,n,m)$ is depicted in Fig.\ \ref{fig4} in function of $\beta$ for some of the
smallest values of $n$ and $m$.
To each curve $\omega^2_{\rm b}(\beta,n,m)$
in Fig.\ \ref{fig4} there corresponds a new family of solutions
bifurcating with the corresponding (i.e.\ of the same values of $\beta,n$) ANO vortex
for $\omega=\omega_{\rm b}(\beta,n,m)$. Based on our numerical results
there is little doubt that for any (integer) value of $n$
a discrete family of curves $\omega^2_{\rm b}(\beta,n,m)$ exist with $m=0,\ldots,n-1$.
The new (static) twisted vortex solutions are therefore expected to exist for
any value of $\beta>1$, and for a given winding number $n$ they comprise a
two parameter set labeled by the twist parameter, $\omega$,
varying continuously, and by the discrete index $m=0,1,\ldots n-1$.
Instead of the twist, $\omega$,
each of the new solution branches can also be parameterized by $f_2^{(m)}$.
For $m=0$, $f_2^{(0)}$ is the value of the condensate at the axis of the twisted vortex.

As can be seen in Fig.\ \ref{fig4}, all curves $\omega^2_{\rm b}(\beta,n,m)$
pass through zero at $\beta=1$. This point corresponds to
the bifurcation of the ``skyrmions'' in the limit of vanishing size parameter $w$
with the corresponding ANO vortices.
The corresponding zero mode solutions of the eigenvalue problem describe
deformations of the ``skyrmions'' with respect to $w$ in the vicinity of $w=0$
\cite{hin}. Our numerics indicate that in the opposite limit, $\beta\to\infty$,
one has $\omega_{\rm b}(\beta,n,m)\to\infty$.

\subsection{New Solutions}
 \begin{figure}[ht]
\hbox to\linewidth{\hss%
  \psfrag{q}{$f_2^{(0)}$}
  \psfrag{E}{${E}/2\pi$}
  \psfrag{b2}{$\beta=2$}
  \psfrag{b4}{$\beta=4$}
  \psfrag{b9}{$\beta=9$}
    \resizebox{8cm}{6.5cm}{\includegraphics{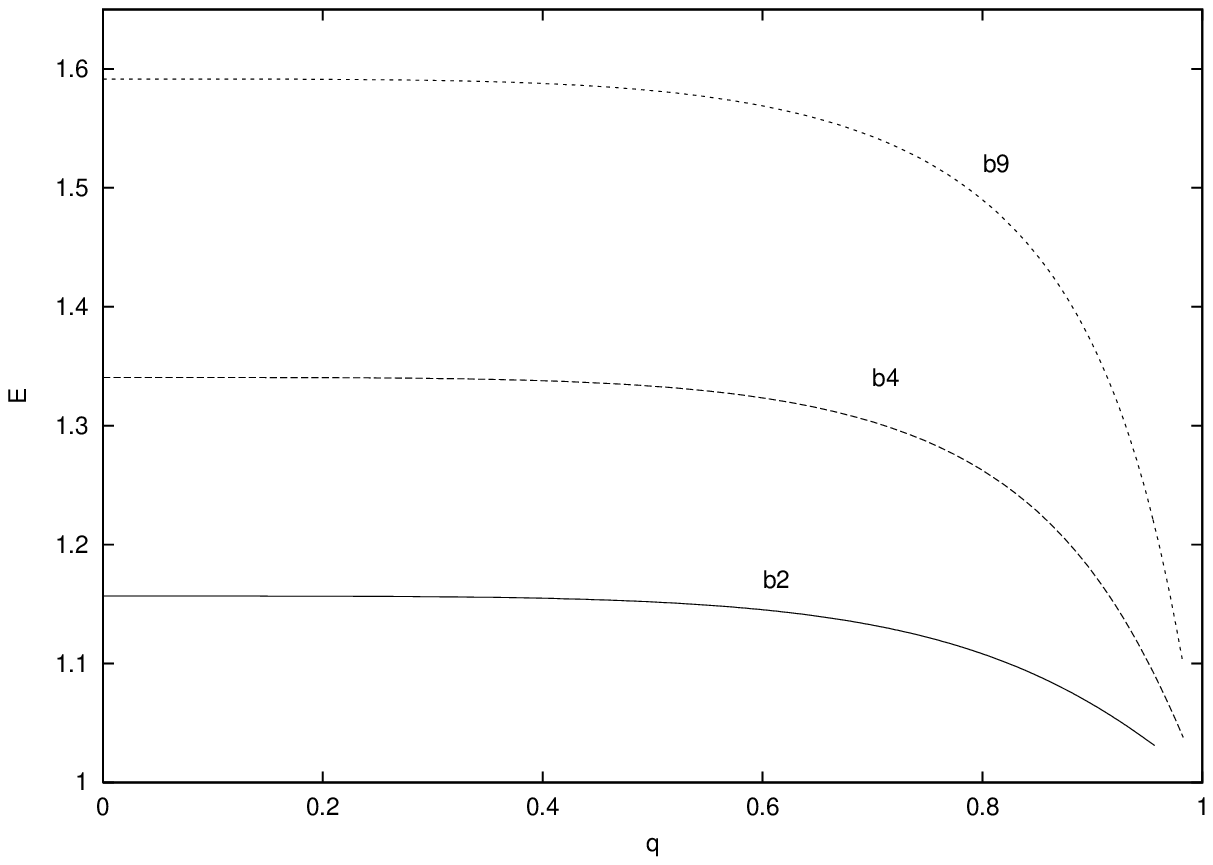}}%
\hspace{5mm}%
  \psfrag{q}{$f_2^{(0)}$}
  \psfrag{I3}{${\cal I}/2\pi$}
  \psfrag{b2}{$\beta=2$}
  \psfrag{b4}{$\beta=4$}
  \psfrag{b9}{$\beta=9$}
    \resizebox{8cm}{6.5cm}{\includegraphics{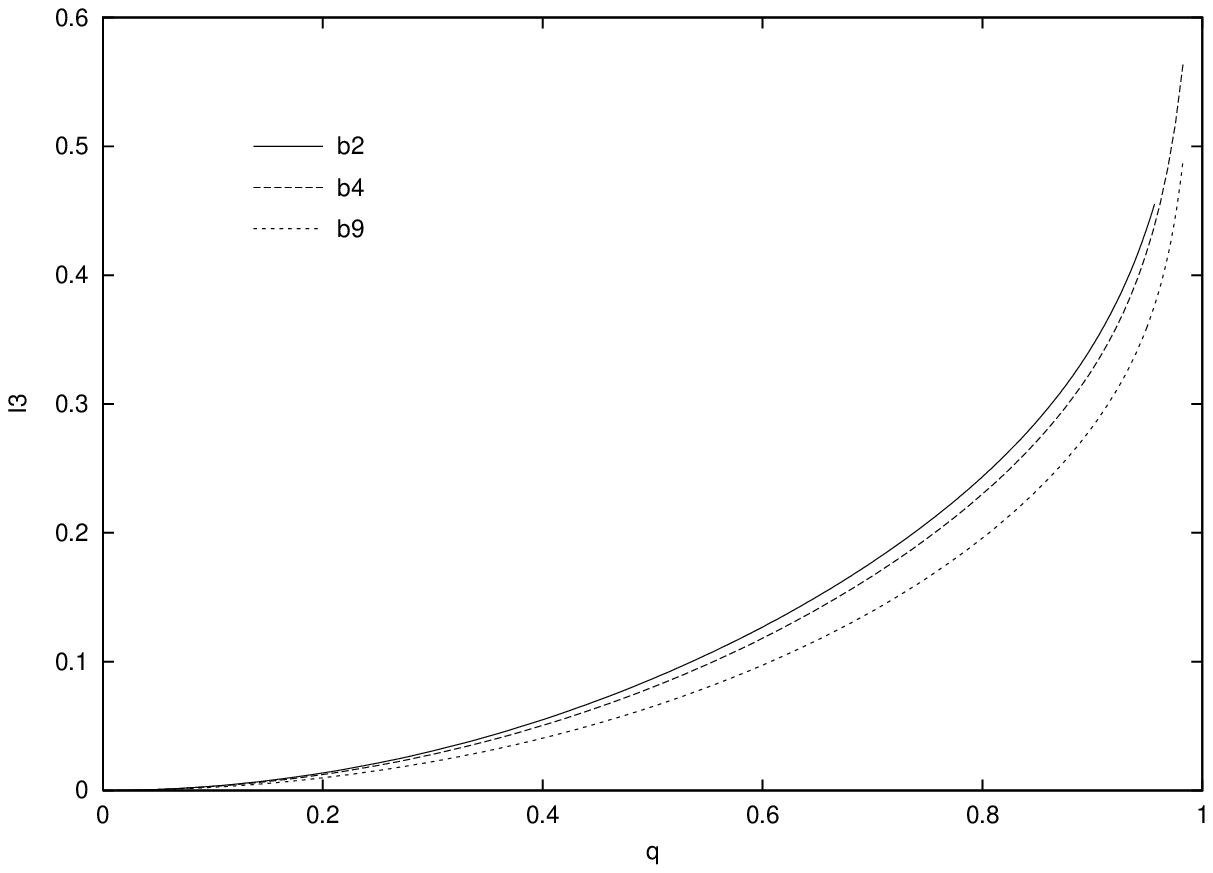}}%
\hss}
\caption{The rest frame energy ${E}/2\pi$ (left) and the rest frame current
${\cal I}/2\pi$ (right) plotted against the value of the condensate $f_2^{(0)}$
at the origin for the solutions with $n=1$ and $\beta=2,\,4,\,9$.}
\label{fig55}
\end{figure}

As exposed in detail in the previous subsection, the known ``magnetic spreading'' instability
of the embedded ANO vortices in the EAH theory indicates in fact the
bifurcation phenomenon, in which new (twisted) solution families branch off from the ANO family.
In Subsection {\bf A} we have already constructed new solutions in the
linear approximation, but
it remains to construct them in the full nonlinear theory.
In view of the mathematical complexity of
Eqs.\ (\ref{cyl-eqs}) we do not have much hope to obtain analytical solutions, although
an existence proof using functional techniques appears to be feasible. In this paper
we present exclusively numerical results, which in our view provide convincing evidence
as to the existence of globally regular twisted solutions.
Adapting the multi shooting procedure described in Ref.\
\cite{FOR} and tested for the ANO vortices (see Table I),
we have succeeded to obtain numerically globally regular solutions of
Eqs.\ (\ref{cyl-eqs}) corresponding to twisted, current carrying vortices.
To obtain the new solutions we start close to the bifurcation point, i.e.\ we choose
values of the shooting parameters close to those of the bifurcating solution, and then we let
vary the value of $\omega$ in the neighborhood of $\omega_{\rm b}^2(\beta,n,m)$
to obtain ``nontrivial'' (i.e.\ different from the ANO vortex) solutions.

In fact for our solutions $\omega$
varies in a finite interval, $0<\omega\leq\omega_{\rm b}(\beta,n,m)$
(for $n>2$ the situation is actually slightly more complicated; see below).
As $\omega$ tends to
$\omega_{\rm b}(\beta,n,m)$,
the functions $a_3\to0$, $f_2\to0$, while
$a\to a_{\scriptscriptstyle\rm ANO}$, $f_1\to f_{\scriptscriptstyle\rm ANO}$, i.e.\
the new class bifurcates with the corresponding ANO vortex. The limit
$\omega\to0$ is somewhat more complicated.
Instead of approaching an untwisted vortex with $\omega=0$ in this limit,
solutions for $\omega\to0$ approach pointwise the exterior part of the
``skyrmions'' in the large size limit ($w\to\infty$), without ever attaining it.
In other words, the point $\omega=0$ does not belong to the phase space
of twisted vortices. This limit will be analyzed in some detail in Subsection
 {\bf E}.
\begin{figure}[ht]
\hbox to\linewidth{\hss%
  \psfrag{X}{$\omega=0.3$}
  \psfrag{x}{$\ln(1+\rho)$}
  \psfrag{f1}{$f_1$}
  \psfrag{f2}{$f_2$}
  \psfrag{a}{$a$}
  \psfrag{a3}{$a_3$}
    \resizebox{8cm}{6.5cm}{\includegraphics{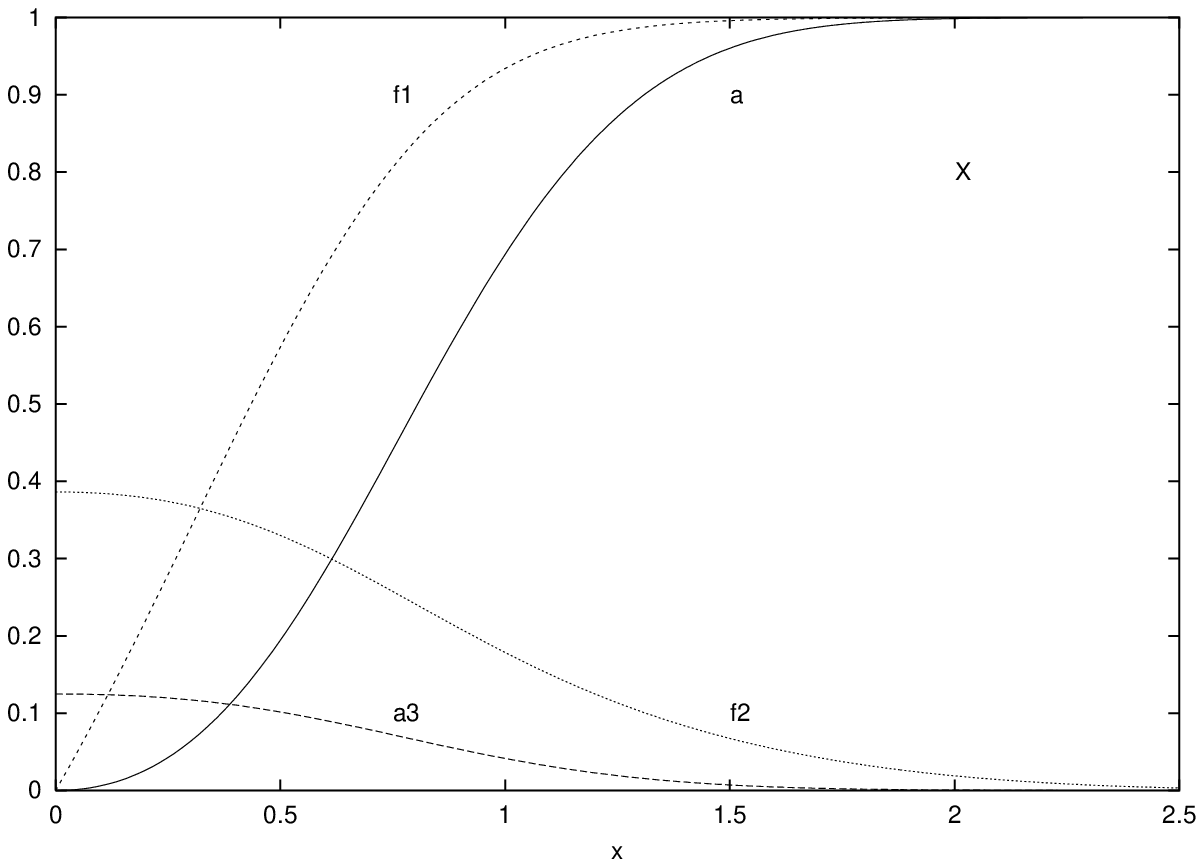}}%
\hspace{5mm}%
  \psfrag{X}{\hspace*{-1cm}$\omega=0.017$}
    \resizebox{8cm}{6.5cm}{\includegraphics{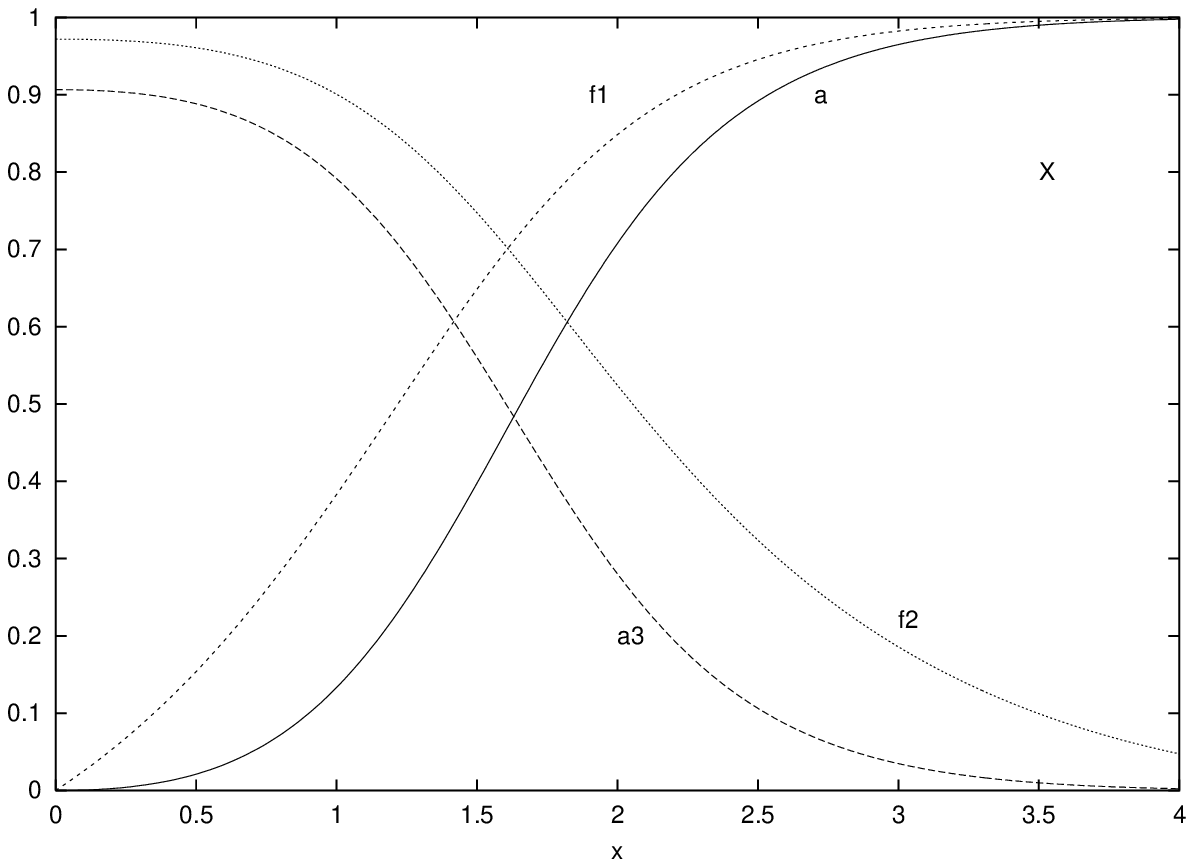}}%
\hss}
\caption{Twisted semilocal vortex solutions for $n=1$, $\beta=2$.}
\label{fig-sols-n=1}
\end{figure}

Every solution of Eqs.\ (\ref{cyl-eqs}) can be Lorentz-boosted, giving rise to a whole
family of vortices, labeled by $\omega_0$, with a given $\omega^2=\omega_3^2-\omega_0^2$.
Passing to the rest frame, where $\omega_0=0$, we observe that
the rest frame energy, $E$,
decreases for decreasing values of $\omega$ or equivalently for increasing values
of the condensate $f_2^{(0)}$
at the origin.
This is illustrated in Fig.\ \ref{fig55}, where ${E}/2\pi$ is
plotted in function of  $f_2^{(0)}$.
As the twist parameter decreases,
the current and the condensate value at the center of the fundamental vortex increase;
see Fig. \ref{fig4}. We have
also plotted in Fig.\ \ref{fig55} the rest frame current, ${\cal I}/2\pi$,
(introduced in
Eq.\  \eqref{restframecurr}) in function of the value of the condensate at the
origin, $f_2^{(0)}$.

For solutions with $n\leq 2$
the decrease of the energy is monotone, so all such twisted vortices
have lower energy than the corresponding ANO ones. For $n>2$ 
the maximal value of the twist parameter can be different from the
bifurcation value. For example for $n=3$ we find that 
$\omega_{\rm max}(\beta,3,0)>\omega_{\rm b}(\beta,3,0)$
(this is recorded in Table IV), and so 
  for a given value of the twist parameter in the interval
$\omega_{\rm b}<\omega<\omega_{\rm max}$
there are two distinct twisted vortices. 
In this case $f_2^{(0)}$ can be chosen instead of $\omega$ to label the solutions, 
since is always increases monotonically as one moves along the solution family. 
For sufficiently large values of $f_2^{(0)}$
both $\omega$ and $E$ start decreasing, and
the energy $E$ of the twisted vortex becomes
smaller than $E_{\rm\scrs ANO}$ while $\omega$ eventually tends to zero. 
The rest frame energy of the twisted vortices is therefore bounded as
\be\label{EEE}
E_{\rm\scrs ANO}(\beta,n)\geq E(\beta,\omega,n,m)> 2\pi n+\omega\,|{\cal I}|,
\end{equation}
for $n=1,2$, while for $n\geq 3$ the first inequality here
is valid when $\omega$ is sufficiently smaller than $\omega_{\rm b} $.
The energy of the new solutions with $n=1,2$ is thus always
{\sl smaller} than that of the ANO vortex.
Therefore we expect the new solutions of winding number $n=1$ with a nonzero current
to be stable.

\begin{figure}[ht]
\hbox to\linewidth{\hss%
  \psfrag{X}{$\omega=0.1$}
  \psfrag{x}{$\ln(1+\rho)$}
  \psfrag{f1}{$f_1$}
  \psfrag{f2}{$f_2$}
  \psfrag{a/n}{$a$}
  \psfrag{a3}{$a_3$}
    \resizebox{8cm}{6.5cm}{\includegraphics{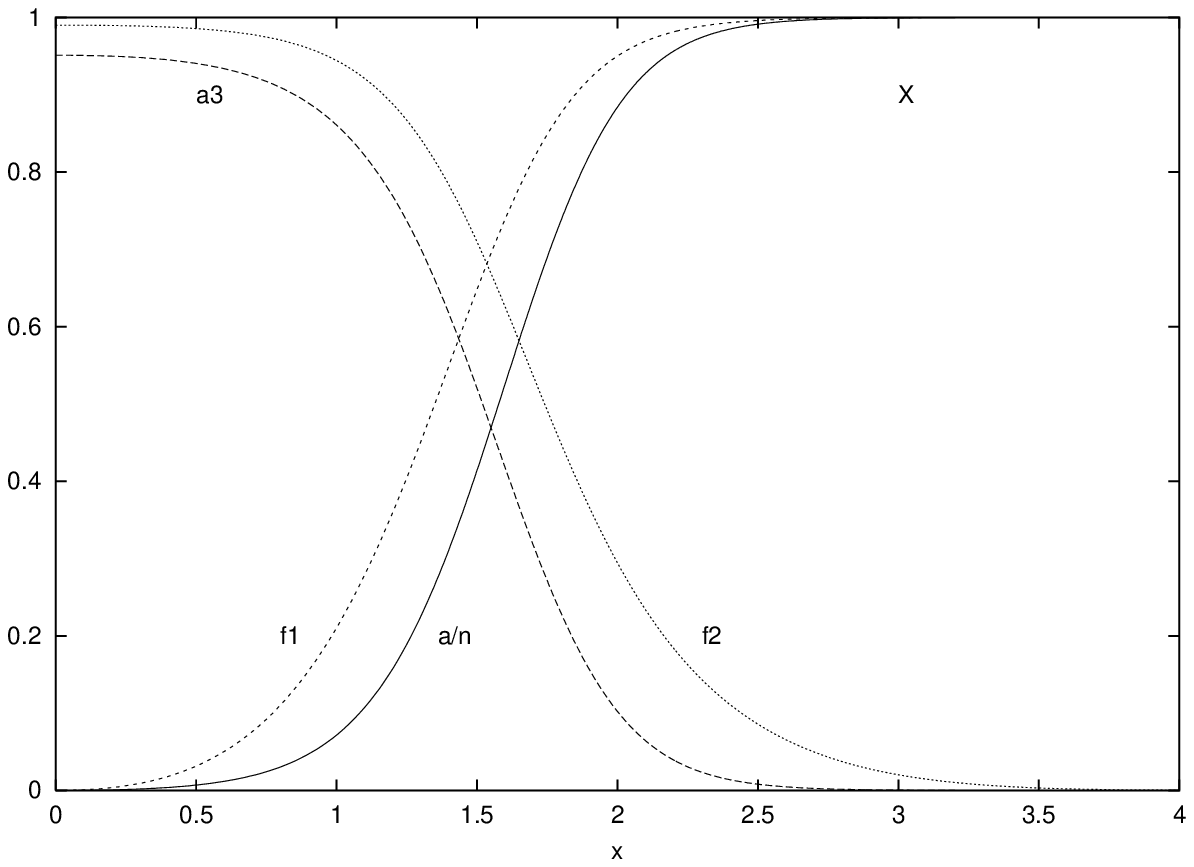}}%
\hspace{5mm}%
\psfrag{f}{$f_1$}
\psfrag{g}{$f_2$}
\psfrag{W}{$a$}
\psfrag{Y}{$a_3$}
  \psfrag{lg}{$\ln(1+\rho)$}
\psfrag{o}{$\omega=0.215$}
    \resizebox{8cm}{6.5cm}{\includegraphics{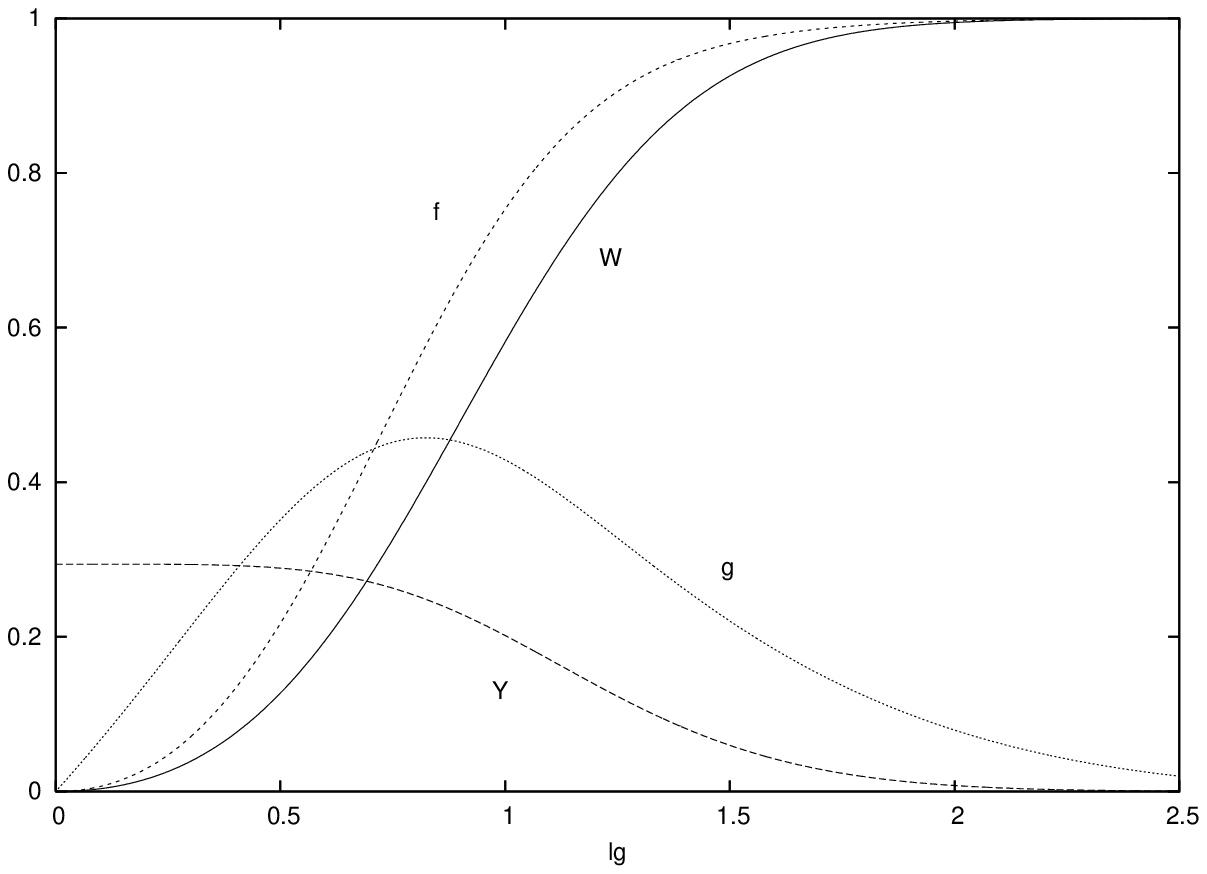}}%
\hss}
\caption{Twisted semilocal vortex solutions for $\beta=2$ and
$n=2,\,m=0,\,\omega=0.1$ (left) and $n=2,\,m=1,\,\omega=0.215$ (right).}
\label{fig-sols-n=2}
\end{figure}
\begin{table}[ht]
\begin{center}
\parbox{15cm}{\caption{Solution parameters for $\beta=2,4,9$ and $n=1$.}}
\vskip 0.3cm
\hspace{10mm}%
\begin{tabular}{c||cccccccc}
$\beta$ & $\omega$ & $f_{\scrs 1}^{\scrs (1)}$ & $f_{\scrs 2}^{\scrs (0)}$ & $a^{\scrs (2)}$ &
$a_{\scrs 3}^{\scrs (0)}$ & ${E}/2\pi$ & ${Q}/2\pi$ & $\mathcal{I}/2\pi$\\
\hline
                & 0.3  & 1.010824 & 0.386212 & 0.546391 & 0.124789 & 1.155180 & 0.170624 & 0.051187\\
                & 0.2  & 0.767653 & 0.705426 & 0.363046 & 0.422267 & 1.131355 & 0.902637 & 0.180527\\
                & 0.1  & 0.534459 & 0.864141 & 0.204976 & 0.663061 & 1.083558 & 3.015082 & 0.301508\\
\raisebox{4.5ex}[0pt]{2}& 0.017& 0.239896 & 0.972041 & 0.051176 & 0.906725 & 1.020255 & 29.91849 & 0.508614\\
\hline
                & 0.6  & 1.305601 & 0.428496 & 0.675653 & 0.107975 & 1.336898 & 0.096934 & 0.058161\\
                & 0.3  & 0.799348 & 0.837937 & 0.354284 & 0.496610 & 1.237417 & 0.869072 & 0.260722\\
                & 0.1  & 0.462641 & 0.947953 & 0.154324 & 0.754081 & 1.106005 & 4.140499 & 0.414050\\
\raisebox{4.5ex}[0pt]{4}& 0.03 & 0.272278 & 0.981800 & 0.062910 & 0.889577 & 1.039866 & 18.60003 & 0.558001\\
\hline
                & 1    & 1.677185 & 0.570981 & 0.808482 & 0.129739 & 1.573669 & 0.087067 & 0.087067\\
                & 0.6  & 1.113827 & 0.849213 & 0.515067 & 0.379056 & 1.443952 & 0.387808 & 0.232685\\
                & 0.3  & 0.734885 & 0.939568 & 0.300118 & 0.589641 & 1.273394 & 1.130746 & 0.339224\\
            & 0.1  & 0.435594 & 0.979182 & 0.136570 & 0.785287 & 1.116579 & 4.663623 & 0.466362\\
\raisebox{4.5ex}[0pt]{9}& 0.08 & 0.394789 & 0.982893 & 0.116465 & 0.812686 & 1.097425 & 6.154640 & 0.492371\\
\hline
\end{tabular}
\end{center}
\end{table}
In Figs.\ \ref{fig-sols-n=1} we depict two sample twisted vortex function profiles for
$\omega=0.3$ and also for a rather small value of the twist, $\omega=0.017$ (with $\beta=2$
and $n=1$).
In Table II we list the relevant parameters of some of the twisted vortex solutions
for $n=1$: the twist, $\omega$, the values of
$f_{\scrs 1}^{\scrs (1)}$, $f_{\scrs 2}^{\scrs (0)}$, $a^{\scrs (2)}$, $a_{\scrs 3}^{\scrs (0)}$
parameterizing the solutions at the origin, the rest frame energy,
${E}$, the invariant charge ${Q}$,
and  the rest frame value of the current,  $\mathcal{I}=\omega Q$.
\begin{table}
\begin{center}
\parbox{15cm}{\caption{Parameters of the fundamental solutions for $\beta=2$ and $n=2$, $m=0$}} \vskip 0.3cm
\begin{tabular}{cccccccc}
$\omega$ & $f_{\scrs 1}^{\scrs (2)}$ & $f_{\scrs 2}^{\scrs (0)}$ & $a^{\scrs (2)}$ & $a_{\scrs
3}^{\scrs (0)}$ & ${E}/2\pi$ & ${Q}/2\pi$ & $\mathcal{I}/2\pi$\\
\hline
0.5  & 0.461753 & 0.743062 & 0.370423 & 0.500163 & 2.389129 & 0.459068 & 0.229534\\
0.1  & 0.075237 & 0.990196 & 0.030212 & 0.951224 & 2.099932 & 4.633715 & 0.463371\\
0.02 & 0.014362 & 0.999592 & 0.001576 & 0.996961 & 2.021527 & 26.21910 & 0.524382\\
\hline
\end{tabular}
\end{center}
\end{table}

\begin{figure}[ht]
\hbox to\linewidth{\hss%
  \psfrag{X}{$E/2\pi$}
  \psfrag{I}{${\cal I}/2\pi$}
  \resizebox{10cm}{6.5cm}{\includegraphics{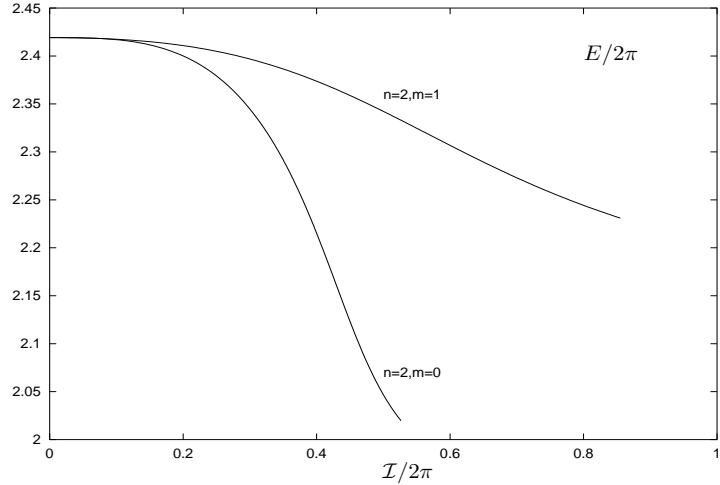}}%
\hss}
\caption{Rest frame energy $E/(2\pi)$ versus the
rest frame current ${\cal I}/(2\pi)$ for two solutions with $n=2$, $m=0,1$ for $\beta=2$.
For $m=0$ and for large currents one has $E/(2\pi)\to2$.}
\label{fig-en-current}
\end{figure}

Next we turn to solutions with  winding number $n>1$. In this case one finds $n$ distinct
one-parameter solutions indexed by $m=0,1,\dots,n-1$, or equivalently by the helicity index
$\nu=1,\ldots, n$ determining the value of the angular momentum in Eq.\eqref{PJ}.
In Figs.\ \ref{fig-sols-n=2} we depict an $n=2, m=0$ twisted vortex with
$\omega=0.1$ and also another one with $n=2, m=1$ and $\omega=0.215$.
For $m>0$ both 
$f_1$ and $f_2$
vanish at the origin, i.e.\ there is no condensate there. While the $\beta=1$
``skyrmions'' with different values of $m$ are all degenerate in energy,
for twisted vortices with $\beta>1$ this degeneracy is lifted.
In fact, solutions with $m>0$ can be viewed as `excitations' of
the `fundamental' solution with $m=0$, since they have higher energy, this is illustrated in
Fig. \ref{fig-en-current}.

\begin{table}[ht]
\begin{center}
\parbox{15cm}{\caption{Parameters of the fundamental solutions for $\beta=2$ and $n=3$, $m=0$}} \vskip 0.3cm
\begin{tabular}{cccccccc}
$\omega$ & $f_{\scrs 1}^{\scrs (3)}$ & $f_{\scrs 2}^{\scrs (0)}$ & $a^{\scrs (2)}$ & $a_{\scrs
3}^{\scrs (0)}$ & ${E}/2\pi$ & ${Q}/2\pi$ & $\mathcal{I}/2\pi$\\
\hline
0.71 & 0.389410 & 0.122714 & 0.647036 & 0.018799 & 3.716103 & 0.010221 & 0.007257\\
0.73 & 0.353581 & 0.380272 & 0.587955 & 0.167458 & 3.717523 & 0.094779 & 0.069189\\
0.4  & 0.064950 & 0.965118 & 0.101566 & 0.878371 & 3.459984 & 1.215289 & 0.486116\\
0.13 & 0.009128 & 0.998709 & 0.007153 & 0.988680 & 3.167331 & 4.627920 & 0.601630\\
\hline
\end{tabular}
\end{center}
\end{table}
\begin{table}
\begin{center}
\parbox{15cm}{\caption{Parameters of the excited solutions for $\beta=2$ and $n=2$, $m=1$}} \vskip 0.3cm
\begin{tabular}{cccccccc}
$\omega$ & $f_{\scrs 1}^{\scrs (2)}$ & $f_{\scrs 2}^{\scrs (1)}$ & $a^{\scrs (2)}$ & $a_{\scrs
3}^{\scrs (0)}$ & ${E}/2\pi$ & ${Q}/2\pi$ & $\mathcal{I}/2\pi$\\
\hline
0.2  & 0.559894 & 0.721888 & 0.649659 & 0.374412 & 2.403820 & 1.288194 & 0.257639\\
0.1  & 0.343013 & 0.976403 & 0.641370 & 0.715442 & 2.321518 & 5.587675 & 0.558768\\
0.04 & 0.212548 & 1.054153 & 0.629574 & 0.873387 & 2.241088 & 20.31268 & 0.812507\\
\hline
\end{tabular}
\end{center}
\end{table}

We tabulate a few values of the pertinent parameters of some fundamental
twisted solutions
with $n=2,3$ in Tables III and IV, and of some excited
ones with $n=2$ and $m=1$ for $\beta=2$ in Table V.

In the limit $\omega\to \omega_{\rm b}(\beta,n,m)$ solutions with different values of $m$
tend to the same ANO vortex, whose energy does not depend on $m$.
In the opposite limit, for small values of the twist and for large currents,
the energy of the fundamental solutions approaches the universal
lower bound $E=2\pi n$ irrespectively of the value of $\beta$,
while for the excited solutions this is not
the case, since their energy grows with $\beta$ and
diverges in the London limit $\beta\to\infty$ for any ${\cal I}$ (for a more detailed
discussion of this limit see the next Subsection).

We now describe some interesting features of the stationary case.
As  has already been discussed stationary solutions can be obtained from
the static ones by applying Lorentz boosts \eqref{boost-omega}.
For static solutions the global charge,
momentum and angular momentum densities are all vanishing,
$j^3_0=T^0_{\phantom0 z}=T^0_{\phantom0 \varphi}=0$. Applying a Lorentz boost 
to a static solution generates
a stationary one whose fields are obtained by simply replacing in
Eq.~\eqref{ansatz2}
$\omega_3=\omega$ and $\omega_0=0$ by
${\omega}_0=-\omega\sinh(\gamma)$, ${\omega}_3=\cosh(\gamma)\omega$.
These stationary solutions have, however,
some different physical properties,
for they have non-vanishing momentum, angular momentum,
and global charge. It is instructive to see how this comes about explicitly.

Let us first remark that the energy momentum tensor of the ANO solutions
is boost invariant, since its only non-zero components
are $T^0_{\phantom0 0}=T^z_{\phantom z z}$, $T^\rho_{\phantom\rho \rho}$,
$T^\varphi_{\phantom\varphi \varphi}$.
When we compute $T^\mu_{\phantom\mu \nu}$ for the
static twisted vortices, we find that $T^0_{\phantom0 \rho}=T^0_{\phantom0 z}
=T^0_{\phantom0 \varphi}=0$,
but otherwise the remaining 7
components are all non-vanishing and describe a complex system of internal
stresses and pressures in the vortex.  The boost invariance of this system
is manifestly broken by non-zero values of $T^0_{\phantom0 0}-T^z_{\phantom z z}$ and
$T^z_{\phantom z \varphi}$, which can be represented in the form
\be                              \label{TTT}
T^0_{\phantom0 0}-T^z_{\phantom z z}=\frac{\omega}{2}\,j^3_3 +\ldots , ~~~~~~~~~
T^z_{\phantom z \varphi}=\frac{\nu}{2}\, j^3_3+\ldots ,
\ee
where dots stand for total derivative terms that vanish upon integration.
If we now apply a Lorentz boost \eqref{boost-omega}, then in view of
the tensorial transformation law, $T_{\mu\nu}\to\tilde{T}_{\mu \nu}=
(\partial x^\alpha/\partial\tilde{x}^\mu)(\partial x^\beta/\partial\tilde{x}^\nu)\,T_{\alpha\beta}$, 
the densities of the energy, momentum and angular momentum
become
\bea\label{tildeeqs}
\tilde{T}^0_{\phantom0 0}&=&T^0_{\phantom0 0} +\sinh^2(\gamma)(T^0_{\phantom0 0}-T^z_{\phantom z z}), \nonumber \\
\tilde{T}^0_z&=& -\sinh(\gamma)\cosh(\gamma)(T^0_{\phantom0 0}-T^z_{\phantom z z}), \nonumber \\
\tilde{T}^0_{\phantom0 \varphi}&=& \sinh(\gamma) T^z_{\phantom z \varphi} \,.
\eea
Integrating Eq.\ \eqref{tildeeqs}  using Eq.\ (\ref{TTT}) and $\int  j^3_3\,d^2 x=4{\cal I}_3=4\omega Q$
one can express the effect of a boost on a static solution as
$E\to\tilde{E}= E+2(\omega_0)^2 Q$, $\tilde{P}=2\omega_0\omega_3Q$,  $\tilde{J}=-2\omega_0\nu Q$,
which completely agrees with Eqs.\ \eqref{energy}, \eqref{PJ} as it should.
Similarly the charge and current densities generated by such a boost are
$\tilde{j}^3_0=-\sinh(\gamma)j^3_3$ and  $\tilde{j}^3_3=\cosh(\gamma)j^3_3$, which gives upon
integration Eq.\ \eqref{III}.

\begin{figure}[ht]
\hbox to\linewidth{\hss%
\hspace{5mm}
  \psfrag{lg}{$\ln(1+\rho)$}
  \psfrag{E}{$T^0_0$}
  \psfrag{Q}{$j^3_3/\omega$}
 \resizebox{8cm}{6.5cm}{\includegraphics{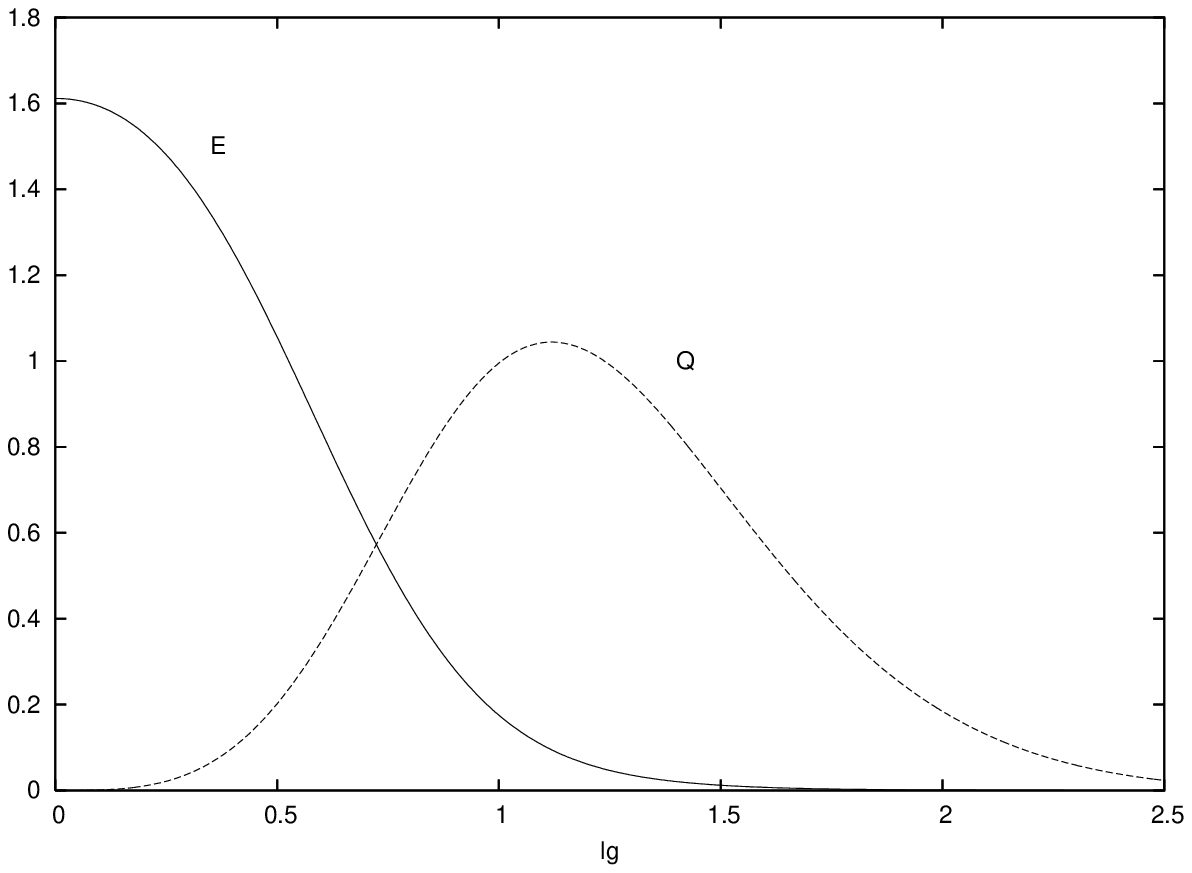}}%
\psfrag{lg}{$\ln(1+\rho)$}
  \psfrag{g0}{$\gamma=0$}
  \psfrag{g2}{$\gamma=2$}
  \psfrag{g3}{$\gamma=3$}
  \resizebox{8cm}{6.5cm}{\includegraphics{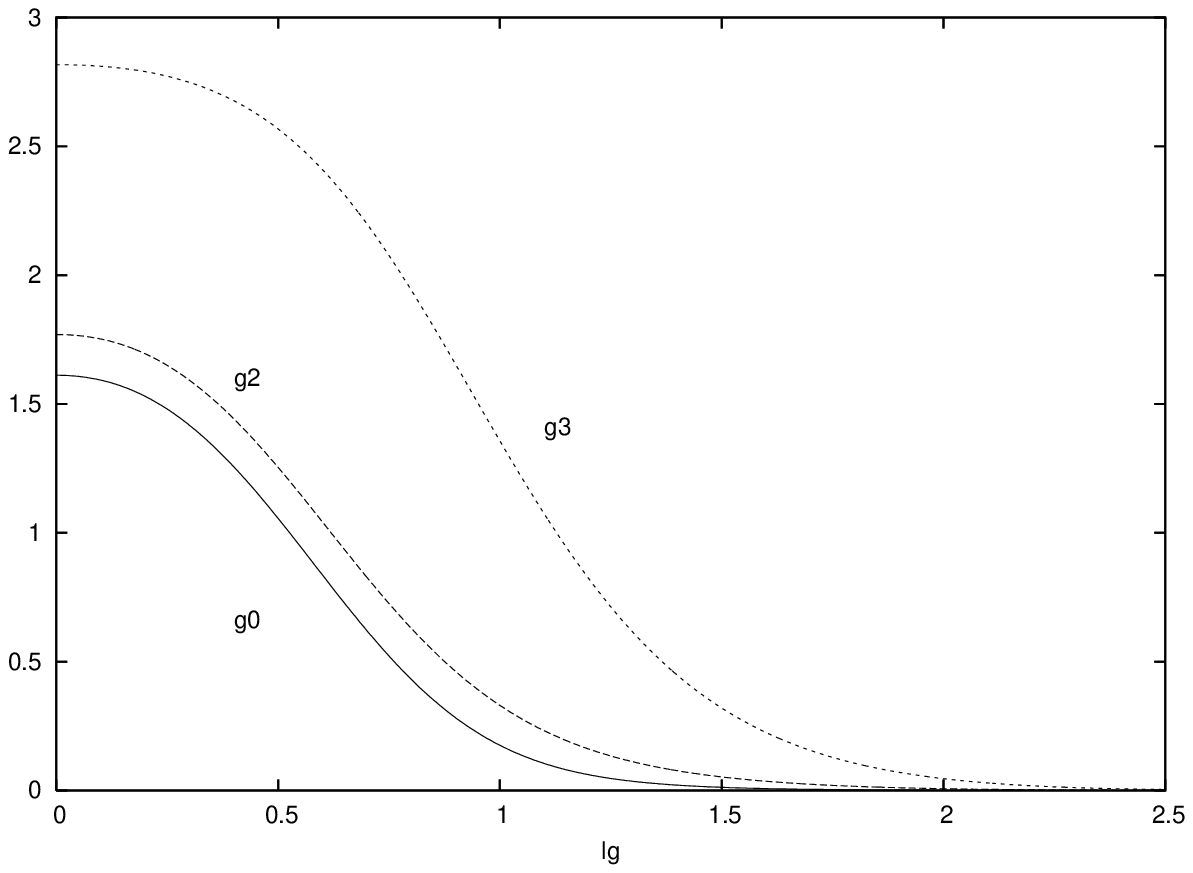}}%
\hss}
\caption{Left: The energy and current densities, $T^0_0$ and $j^3_3$,
for a static solution with $\beta=2$, $\omega=0.19$, and $n=1$.
Right:
Energy densities, $T^0_0$,  of the corresponding stationary solutions
for some values of the boost parameter, $\gamma$.
}
\label{fig-enrexc}
\end{figure}

Summarizing, twisted vortices are not Lorentz boost invariant.
The densities of the energy, momentum,
angular momentum, global charge and the current of a boosted static solution
are given by linear combinations of $T^0_{\phantom0 0}$ and $j^3_3$, 
up to terms that vanish upon integration
(see Fig.\ref{fig-enrexc}).
These stationary solutions can be
labelled by the value of the boost parameter, $\gamma$, or equivalently by $\omega_0$, or
equivalently by their momentum $P$.

\subsection{The $\beta\to\infty$ limit}
\begin{figure}[ht]
\hbox to\linewidth{\hss%
  \psfrag{X}{\hspace*{1.5cm}$\omega=1$}
  \psfrag{x}{$\ln(1+\rho)$}
  \psfrag{f1}{$f_1$}
  \psfrag{f2}{$f_2$}
  \psfrag{a}{$a$}
  \psfrag{a3}{$a_3$}
    \resizebox{8cm}{6.5cm}{\includegraphics{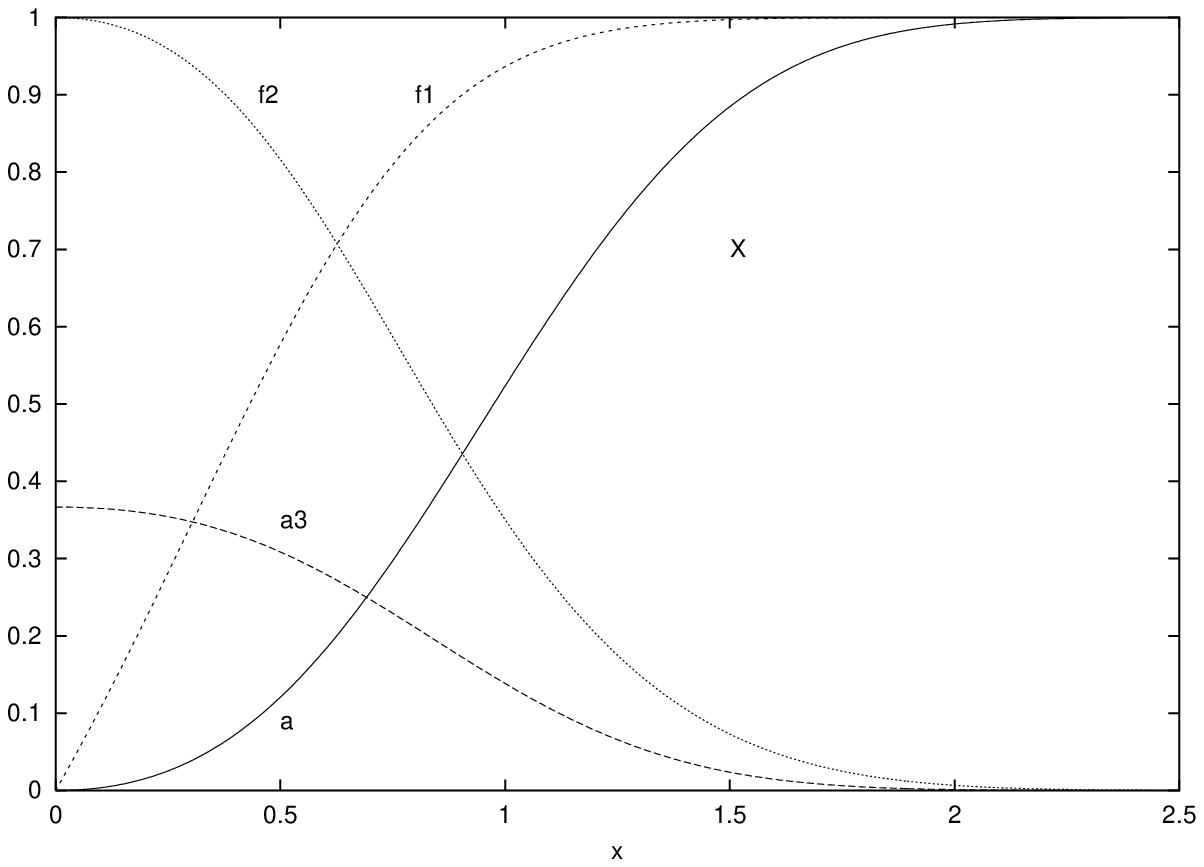}}%
\hspace{5mm}%
  \psfrag{X}{$\omega=0.01$}
  \psfrag{x}{$\ln(1+\rho)$}
  \psfrag{f1}{$f_1$}
  \psfrag{f2}{$f_2$}
  \psfrag{a}{$a$}
  \psfrag{a3}{$a_3$}
    \resizebox{8cm}{6.5cm}{\includegraphics{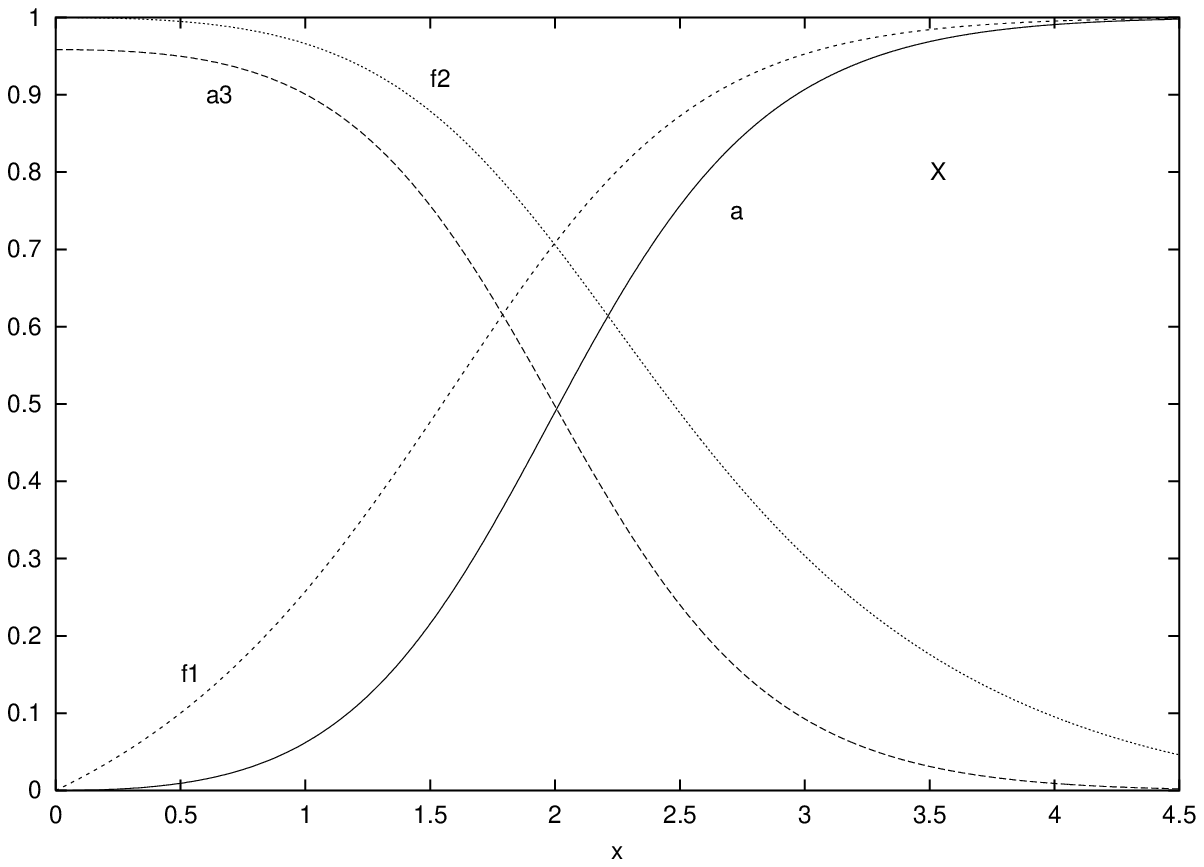}}%
\hss}
\caption{Semi-local vortex solutions for $n=1$, $\beta=\infty$.}
\label{fig-inf}
\end{figure}

It is instructive to discuss in some detail the $\beta=\infty$ limiting theory,
in which case the scalar fields are constrained as
\be\label{constraint}
\Phi^\dagger\Phi=|f_1|^2+|f_2|^2\equiv1,
\end{equation}
when the SU(2) EAH model reduces to the gauged $\mathbf{CP}^1$-model.
It is convenient to parameterize the scalar fields as:
\be
f_1=\cos\theta\,,\quad f_2=\sin\theta\,,
\end{equation}
 and then the field equations can be written as:
\begin{subequations}\label{betainfeqs}
\begin{align}
\frac{1}{\rho}{(\rho{a_3}')}'&=2[a_3-\cos^2\theta]\,,\\
 \rho{\biggl(\frac{a'}{\rho}\biggr)}'&=2\biggl[a-\bigl(1-\frac{\nu}{n}\cos^2\theta\bigr)\biggr],\\
\frac{1}{\rho}(\rho{\theta}')'&=\frac{1}{2}\left[\omega^2(2a_3-1)-
\frac{n\nu}{\rho^2}\biggl(2a-\frac{n+m}{n}\biggl)\right]\sin(2\theta)\,.
\end{align}
\end{subequations}
The local behaviour of the fields near $\rho=0$ is not analytic in $\rho$,
it is given by
\be\label{betainf-ori}
a=(a^{(2)}-\frac{m}{n}\ln\rho)\rho^2+\dots\,,\quad a_3=a_3^{(0)}+\ldots\,,
\quad\theta=c\rho^{\sqrt{n^2-m^2}}+\ldots\,.
\end{equation}
For large values of $\rho$ the asymptotic series of the function $\theta$
is easily seen to be
\be
\theta=\pi/2-D\frac{e^{-\omega\rho}}{\sqrt{\rho}}+\ldots\,,
\end{equation}
while
the various possible asymptotic behaviours of $a$, $a_3$ for $\rho\to\infty$
depending on $\omega\gtrless1/\sqrt{2}$, are still given by the formulae (\ref{inf}a,b).
The profile functions of the current carrying vortex solution of Eqs.\ (\ref{betainfeqs})
are qualitatively similar to those of Eqs.\ (\ref{cyl-eqs}), for a sample profiles see
Figs.\ \ref{fig-inf}.
In Table VI we give the relevant parameters of some solutions with winding number $n=1$
for various values of the twist, $\omega$.
\begin{table}[ht]
\begin{center}
\parbox{15cm}{\caption{Parameters of the $n=1$, $\beta=\infty$ solutions } }\vskip 0.3cm
\begin{tabular}{ccccccc}
$\omega$ & $f_{\scrs 1}^{\scrs (1)}$ & $a^{\scrs (2)}$ & $a_{\scrs 3}^{\scrs (0)}$ & ${E}/2\pi$ &
${Q}/2\pi$ & $\mathcal{I}/2\pi$\\
\hline
 5    & 6.030722 & 1.812331 & 0.064034 & 2.991526 & 0.018523 & 0.092613\\
 1    & 1.442394 & 0.640988 & 0.366642 & 1.753574 & 0.274199 & 0.274199\\
 0.1  & 0.417767 & 0.125292 & 0.804925 & 1.124363 & 5.046921 & 0.504692\\
 0.01 & 0.155052 & 0.022236 & 0.958536 & 1.017715 & 78.02346 & 0.780235\\
\hline
\end{tabular}
\end{center}
\end{table}
For any winding number $n$ the energy of the fundamental $(m=0)$ $\beta=\infty$ solutions remains
bounded for any finite value of $\omega$.
We emphasize that our numerical evidence does suggest $\omega_{\rm b}(\infty)=\infty$.
i.e.\ the parameter $\omega$ varies in the interval $(0,\infty)$.

On the other hand, the energy of the excited $(m>0)$ $\beta=\infty$ solutions
diverges, since the $(m-na)^2f_2^2/\rho$ term in the energy
density \eqref{energy} causes a logarithmic divergence in the total energy as $\rho\to0$.
This behaviour is very similar to the case of the $\beta\to\infty$ limiting ANO vortex, whose energy also
diverges logarithmically.
\subsection{Qualitative description of the phase space}
\begin{figure}[h]
    \psfrag{q}{$f_{\scrs 2}^{\scrs (0)}$}
    \psfrag{beta}{$\beta$}
    \psfrag{sigma}{$\omega$}
    \psfrag{b1}{\hspace{-0.3cm}$\beta=1$}
    \psfrag{b2}{$\beta=2$}
    \psfrag{b4}{$\beta=4$}
    \psfrag{b9}{$\beta=9$}
    \psfrag{wbif}{$\omega_{\rm b}(\beta)$}
    \resizebox{18cm}{14cm}{\includegraphics{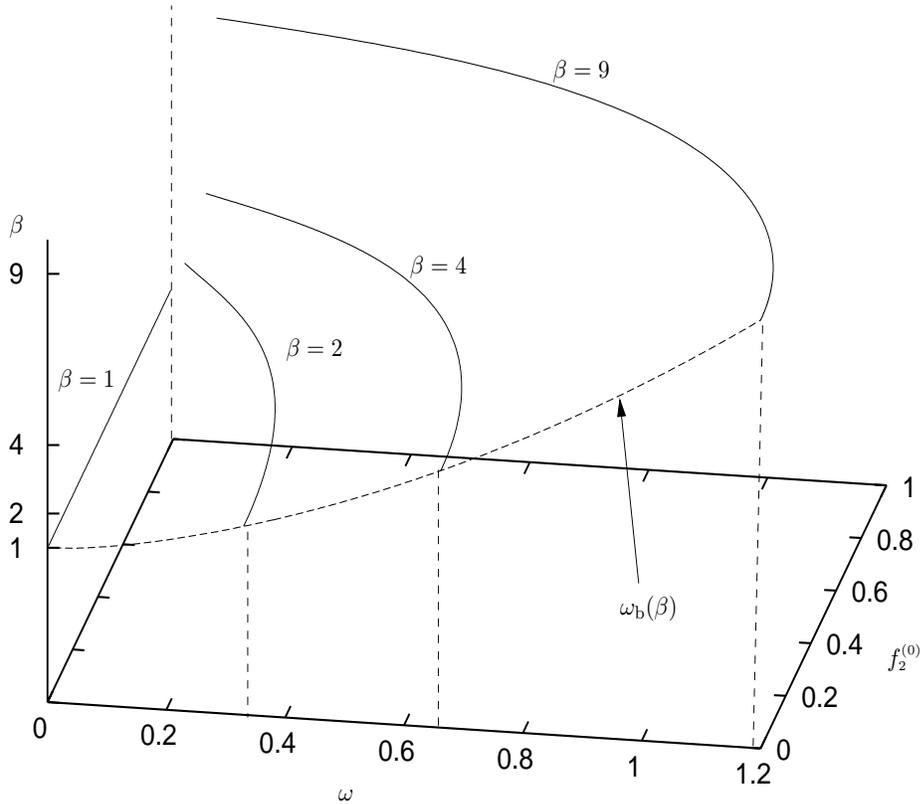}}%
\vskip-0.3cm
    \caption{Phase space of the fundamental $n=1$ twisted vortices.}
\label{fig-phasespace}
\end{figure}

The phase space of the fundamental ($m=0$) twisted vortex solutions with winding number
$n=1$ is depicted in Fig.\ \ref{fig-phasespace},
using the explicit parameters appearing in equations \eqref{cyl-eqs}, $\beta$, $\omega$,
and the magnitude of the condensate at the origin, $f_2^{(0)}$, as coordinates.
%
In Fig.\ \ref{fig-phasespace} a one parameter family of twisted vortices
for a fixed value of $\beta$ corresponds to a curve in the
($f_2^{(0)}$, $\omega$) plane.
The ANO vortex for some fixed value of $\beta$ corresponds to a $\beta=$const.\ straight
line in the $f_2^{(0)}=0$ plane.
The curve $\omega_{\rm b}(\beta)$ is also indicated in the $f_2^{(0)}=0$ plane,
where the bifurcation of the new solutions with the ANO vortices takes place.
Three families of twisted solutions have been indicated as solid curves
for the values $\beta=2,4,9$, together with the ``skyrmion'' solutions satisfying
the \bogo\ equations ($\beta=1$). The curves terminate at the smallest values of $\omega$
we could compute. They should all approach indefinitely the
$(\omega=0\,,f_2^{(0)}=1)$ line,
without ever reaching it.

\begin{figure}[h]
    \psfrag{q}{$f_{\scrs 2}^{\scrs (0)}$}
    \psfrag{E}{${E}/2\pi$}
    \psfrag{sigma}{$\omega$}
    \psfrag{b1}{$\beta=1$}
    \psfrag{b2}{$\beta=2$}
    \psfrag{b4}{$\beta=4$}
    \psfrag{b9}{$\beta=9$}
    \psfrag{binf}{$\beta=\infty$}
    \psfrag{Eno}{${E}_{\scriptscriptstyle ANO}/2\pi$}
    \psfrag{approximateCP1lump}{approximate $\mathbf{CP}^{\scr 1}$ lump}
    \resizebox{18cm}{15cm}{\includegraphics{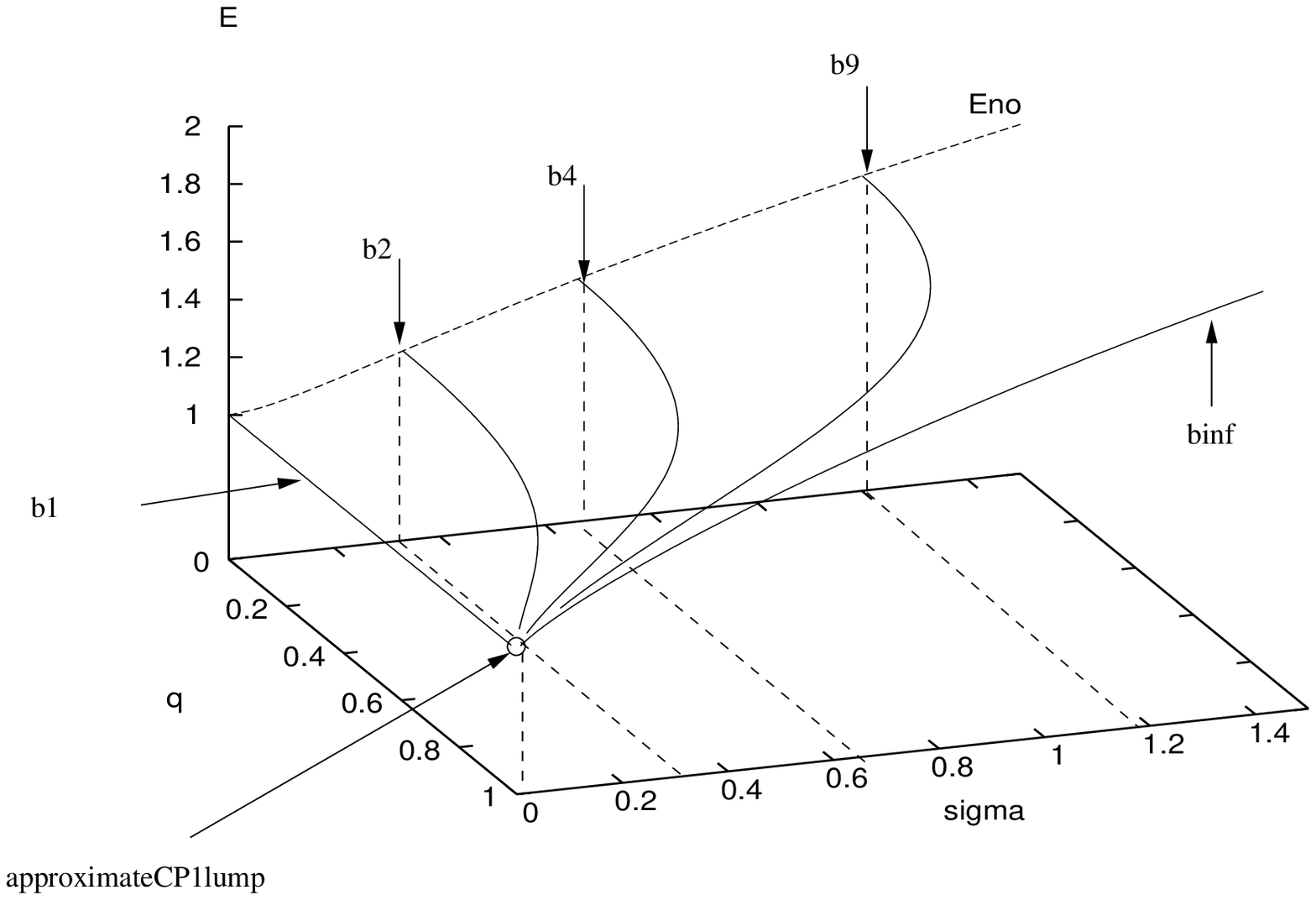}}%
    \vskip-0.1cm
    \caption{Energy of the fundamental twisted vortices.}
     \label{fig-energyspace}
\end{figure}

We have also depicted the ``energy landscape'' of the fundamental solutions
with winding number $n=1$ in Fig.\ \ref{fig-energyspace}.
%
In Fig.\ \ref{fig-energyspace} a family of solutions for a fixed value of
$\beta$ corresponds to a solid curve in the
($f_2^{(0)}$, $\omega$) plane, and we have also indicated the corresponding ANO vortex,
located at the endpoint of the curve in the $f_2^{(0)}=0$ plane where
the new family bifurcates with it.
The family of solutions for $\beta=\infty$ is represented
by the solid curve in the $f_2^{(0)}=1$ plane.
It is clearly seen that as the value of $\omega$ decreases
from $\omega_{\rm b}(\beta)$ towards $\omega=0$,
the energy of all twisted vortices decreases towards
the \bogo\ bound, $E=2\pi $, {\sl without ever attaining it},
irrespectively of the value of $\beta$.
An important point is that solutions with nonzero values of $\omega$ do not have a smooth
$\omega\to0$ limit, i.e.\ the point $(f_2^{(0)}=1\,,\omega=0)$ corresponding
to the approximate $\mathbf{CP}^{\scr 1}$ lump does not belong
to the phase space as is indicated on Fig.\ \ref{fig-energyspace} by a circle.

\subsection{The zero twist limit}

As we have already remarked, the limit of zero twist, $\omega\to0$ is rather
nontrivial. After a brief description of the behaviour of the
solutions for small values of $\omega$,  we shall give a heuristic
explanation of the observed behaviour without any attempt of mathematical rigor.

\begin{figure}[ht]
\hbox to\linewidth{\hss%
  \psfrag{x}{$\omega$}
  \psfrag{binf}{$\beta=\infty$}
  \psfrag{b9}{$\beta=9$}
  \psfrag{b4}{$\beta=4$}
  \psfrag{b2}{$\beta=2$}
    \resizebox{8cm}{6.5cm}{\includegraphics{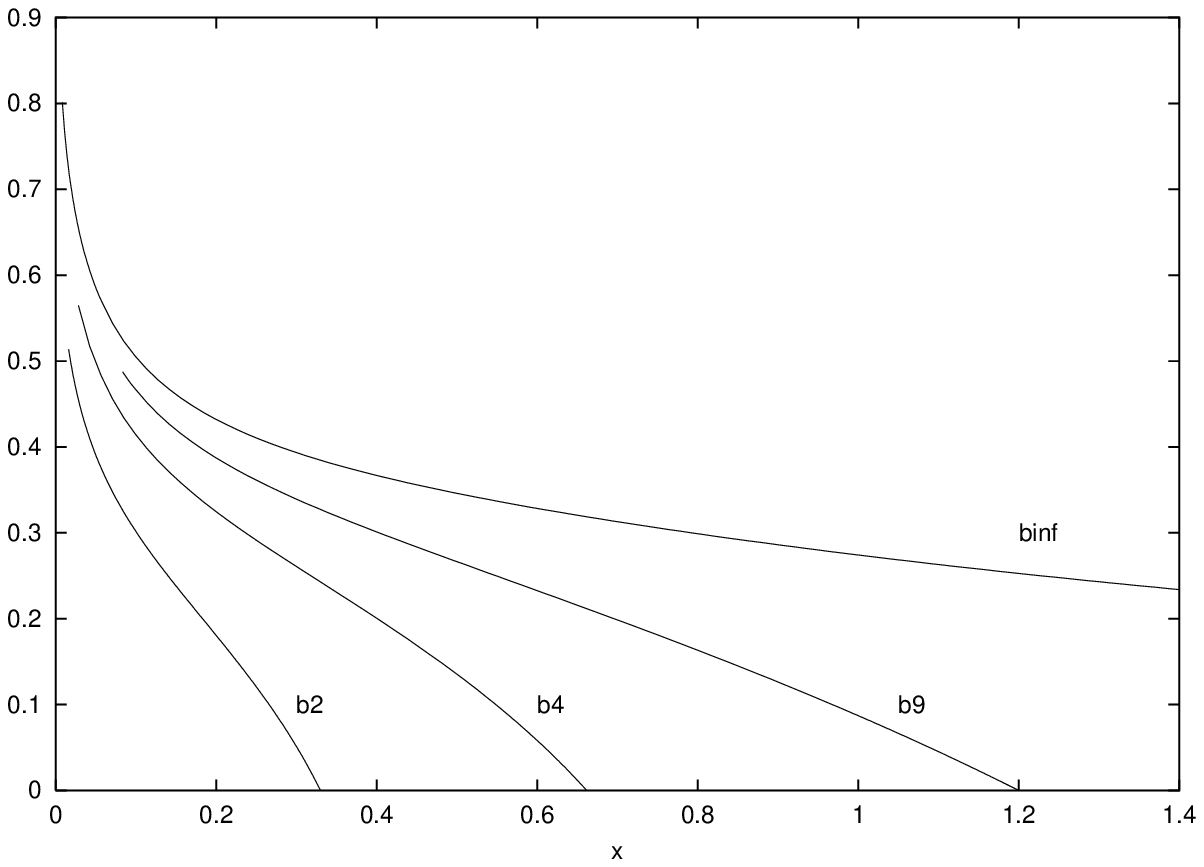}}%
\hspace{5mm}%
  \psfrag{x}{$\mathcal{I}/2\pi$}
  \psfrag{n1}{$n=1$}
  \psfrag{n2}{$n=2$}
  \psfrag{n3}{$n=3$}
    \resizebox{8cm}{6.5cm}{\includegraphics{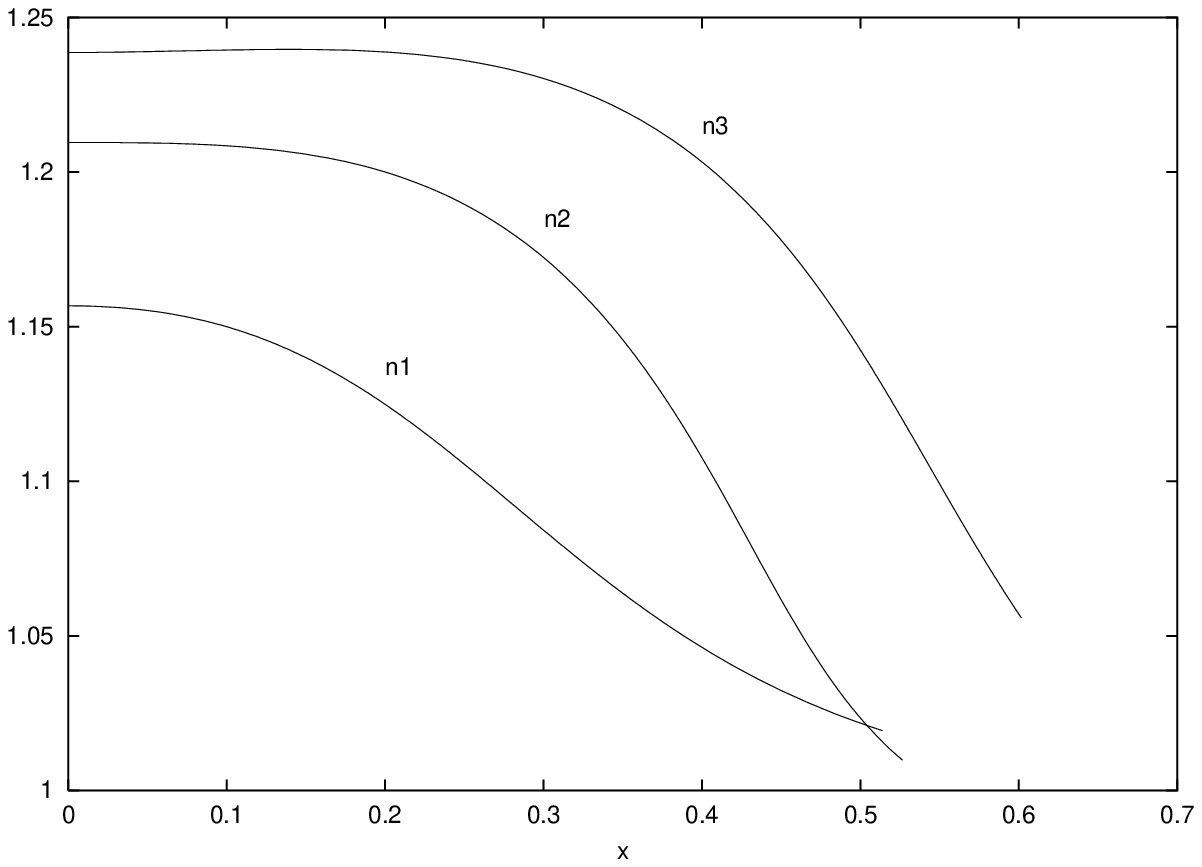}}%
\hss}
\caption{Left : the current, $\mathcal{I}/2\pi$,
as a function of $\omega$ for $n=1$, $\beta=2,4,9,\infty$.
Right : The energy, ${E}/(2\pi n)$, against the current for $\beta=2$ and $n=1,2,3$.}
\label{fig-current}
\end{figure}

One observes that
when the twist parameter $\omega$ decreases with fixed $\beta, n,m$,
the solutions behave differently in a region containing the origin (interior region)
and in the corresponding outside region.
In the outside region the asymptotic behaviour of the solutions
is dominated by the $f_2$ amplitude due to its slowest falloff, but what is more,
the solutions
approach more and more the ``skyrmion'' configurations described in Section VI.
At the same time the value of the current is getting larger and larger,
in fact for $n=1$ we expect
that it will grow without bounds. We depicted the current in
function of the twist parameter
for $n=1$ in Fig.\ \ref{fig-current}.
Irrespectively of the value of $\beta$ the energy of the fundamental solutions ($m=0$)
approaches a universal lower bound which is nothing but the \bogo\ lower bound, i.e.\
for $\omega\to0$ we observe that $\mathcal{I}$ is growing without any apparent limit and
\be
E\to 2\pi n\,,\quad f_2^{(0)}\to 1\,.
\end{equation}
The excited solutions with $m\ne0$ also approach
the same limiting ``skyrmion'' configurations \eqref{skyrm-limit}
depending only on $\nu$.
Unfortunately,
with our numerical procedure we could not obtain solutions for values
of $\omega$ smaller than
$\approx10^{-2}$.
Nevertheless these numerical results enable us to give a
heuristic description of the $\omega\to0$
limit.
It turns out that the solutions behave in this
limit essentially in the same way as the large size ``skyrmions''.
Let us introduce a new function $C(\rho)$ as
\be\label{CCC}
f_2(\rho)=x^{-\nu}\,C(\rho) f_1(\rho),
\end{equation}
where $C(0)=1$, $x=\rho/w$, and $w$ is the ``size'' parameter.
Similarly to the skyrmion case, we will now study the limit
$w\to\infty$. With
\be
s=\frac{1}{w^2}
\end{equation}
the field equations \eqref{cyl-eqs} read
\begin{subequations}\label{lim-eqs}
\begin{align}
s\,\frac{1}{x}\,(x a_3')' =2a_3(f_1^2+f_2^2)-2f_2^2& \,,\\
 s\,x{\biggl(\frac{a'}{x}\biggr)}' = 2f_1^2(a-1)+2f_2^2(a-\frac{m}{n}) &\,,\\
 s\,\biggl\{\frac{1}{x}(x f_1')' - n^2\frac{(1-a)^2}{x^2}f_1\biggr\} =
f_1\left[\omega^2a_3^2-
\beta(1-f_1^2-f_2^2)\right]&\,,\\
\left\{C''+\left(2\frac{{f_1}'}{f_1}+\frac{1-2\nu}{x}\right)C'
-\frac{2\nu}{x}\left(\frac{{f_1}'}{f_1}-\frac{n(1-a)}{x}\right)C\right\} =
\frac{\omega^2}{s}(1-2a_3)C&\,,
\end{align}
\end{subequations}
where now $'$ denotes the derivative with respect to $x$.
Assuming that
\be
\lim_{s\to 0}\omega^2=0,
\end{equation}
and that the derivatives remain bounded in an interval, $\mathcal D$,
equations (\ref{lim-eqs}a-c) imply on $\mathcal D$ for $s\to 0$
\be
0=(f_1^2+f_2^2)a_3-f_2^2\,,\quad 0=f_1^2(a-1)+f_2^2\left(a-\frac{m}{n}\right)\,,\quad
0=\beta(f_1^2+f_2^2-1)f_1,
\end{equation}
whose solution can be parameterized with a single function $\Theta(x)$:
\be                                     \label{00}
f_1=\sin( \Theta)\,,\quad f_2=\cos(\Theta)\,,\quad
a_3=\cos^2(\Theta)\,,\quad n(1-a)=\nu\cos^2(\Theta).
\end{equation}
The function $C(x)$ can now be expressed as
\be\label{C-expr}
C=x^{\nu}\,\frac{f_2}{f_1}=x^{\nu}\cot(\Theta).
\end{equation}
Inserting \eqref{C-expr} to the equation for $C(x)$ (\ref{lim-eqs}d)
one obtains
\be\label{theta-eq}
\Theta''+\frac{1}{x}\,\Theta'=\frac14\,(\lambda+\frac{{\nu}^2}{x^2})\sin (4\Theta)\,,
\end{equation}
where $\lambda\equiv\lim_{s\to 0}{\omega^2}/{s}$.
Since the boundary conditions, $\Theta(x)\to\pi/2$ for $x\to\infty$ and
$\Theta(x)\to0$ as $x\to0$, ensure that the fields $f_1, f_2, a, a_3$ have the
{\sl qualitatively} correct {\sl global} asymptotic behavior not only on $\mathcal D$
but by extension on the whole $0\leq x<\infty$,
we seek solutions of \eqref{theta-eq}
subject to these boundary conditions.
There are now two possible cases: either $\lambda$ is unbounded or it is finite.
In fact  $\lambda$ must be bounded, since otherwise $\Theta(x)$ could not
satisfy the boundary conditions.
A simple positivity argument shows that, unless $\lambda=0$, equation \eqref{theta-eq}
does not admit solutions satisfying the boundary conditions.
When $\lambda=0$, the solution is given by
\be\label{lim}
\sin(\Theta)=\frac{x^{\nu}}{\sqrt{1+x^{2\nu}}}.
\end{equation}
We therefore conclude that in the $1/\sqrt{s}=w\to\infty$ limit
one has
$\lim_{s\to 0}\omega^2/s=0$,
and the solutions approach the following configuration 
\be\label{CP1}
f_1=\frac{x^{\nu}}{\sqrt{1+x^{2\nu}}}\,,\quad
f_2=\frac{1}{\sqrt{1+x^{2\nu}}}\,,\quad
n(1-a)=\frac{\nu}{{1+x^{2\nu}}}\,,\quad
a_3=\frac{1}{{1+x^{2\nu}}}\,,
\end{equation}
and furthermore  $C(x)\equiv1$.
This limiting configuration is indeed the same as the one obtained in
the large size skyrmion
limit \eqref{skyrm-limit}.
What is interesting, is that this limiting behavior is {\sl universal}, i.e.\
it does not depend on $\beta$.
Similarly to the skyrmions, the limiting configurations depend only on $\nu$,
and so solutions with $n=\nu,\nu+1,\ldots$ have the same limit.
Only the fundamental $m=0$ solutions approach the limiting skyrmion globally, since they
have the same boundary conditions at the origin as (\ref{CP1}), and so their energy
approaches in the limit the \bogo\ bound, i.e.\ $E=2\pi n $.
 The excited solutions ($m\ne0$) have
different boundary conditions at $x=0$, and they approach
the limiting configuration in a domain not containing the origin.

Finally we present a simple virial argument which shows that at least
in the $\beta=\infty$ theory there are no nontrivial ``skyrmion'' solutions.
From the energy functional \eqref{energy} one can obtain the following virial relation
for static configurations:
\be\label{virial}
\int_0^\infty d\rho\,n^2\frac{{a'}^2}{\rho}=\int_0^\infty \rho d\rho\,\biggl\{
2\omega^2\bigl[ f_1^2a_3^2+f_2^2(1-a_3)^2\bigr]
+\beta\left(1-|f|^2\right)^2 \biggr\}\,,
\end{equation}
which immediately implies that for $\omega=0$ the $\beta=\infty$ theory,
when the the potential energy term $\propto\beta$ is absent,
{\sl does not} admit nontrivial finite energy solutions.
Although for finite values of $\beta$, the above simple virial argument does
not exclude the possible existence of ``skyrmions'' for $\omega=0$,
the analysis of the $\omega\to0$ limit
has indicated that in fact irrespectively of the value of $\beta$,
$f_1^2+f_2^2\to1$, therefore it is unlikely that nontrivial ``skyrmions''
would exist apart from the known ones for $\beta=1$.

\section{Summary and concluding remarks}
\setcounter{equation}{0}
In summary, we have presented a new class of straight, twisted string solutions of the
SU(2)
symmetric semilocal Abelian Higgs model.
These twisted strings have finite charge, energy, momentum, and angular
momentum per unit length, and they also carry a global current.
The solutions are not Lorentz boost invariant in the $t,z$ plane,
and there exists a rest frame where their charge, momentum, and angular momentum vanish.
The current does not vanish, however, and so it is an
essential physical parameter.

The twisted solutions exist for $\beta>1$ and they can be labeled by the frequency $\omega_0$
and twist $\omega_3$, by the winding number
$n=1,2,\ldots$, and by an additional integer, $m=0,1,\ldots n-1$.
Related to these, there are four independent physical
parameters of the vortex: the momentum $P\sim\omega_0$, current ${\cal I}\sim\omega_3$,
magnetic flux $\sim n$, and angular momentum $J\sim (n-m)$.
For ${\cal I}\to 0$ the twisted solutions bifurcate with the embedded ANO
vortices. The current seems to range in the infinite interval, at least for $n=1$,
and for $|{\cal I|}\to\infty$ the solutions approach the
$\mathbf{CP}^{\scr 1}$~lump(s) just as the ''skyrmions'' do in the large size limit.
The energy in the rest frame {\it decreases} as $|{\cal I}|$ grows
and, if $m=0$, it tends to
the absolute lower bound $2\pi n $ in the zero twist limit.
The fundamental solutions with $m=0$ have the lowest energy
for a given $n$ and ${\cal I}\neq 0$. They have finite energy also in the
London limit $\beta\to\infty$, whereas the energy of the excited strings
with $m\ne0$ diverges.
As $\omega\to 0$, the solutions approach the
limiting ``skyrmion'' configurations.

Perhaps the most important property of the twisted strings is the fact that
their energy is lower as compared
to that of the ANO vortices.
They are therefore energetically more favorable to be produced during phase
transitions, and so one would expect that the formation of such twisted strings
is an important issue.
This can be called `spontaneous superconductivity'
and this may imply interesting applications in cosmological models.
The new twisted vortices could also play a role in solid state physics in models
of various superconductors with several order parameters
(e.g. two-band superconductivity models
containing a vector and two complex scalar fields \cite{Sigrist,Babaev}).
In superfluid $^3$He twisted vortices qualitatively similar to ours
(Salomaa-Volovik vortices) have been experimentally observed \cite{Salomaa}.
From the point of view of high energy physics the twisted semilocal strings
should have their twisted $Z$-strings counterparts in the bosonic sector of Standard Model
of Electroweak interactions.

An important property of the superconducting semilocal strings it that, similarly
to the topological ANO vortices, they cannot terminate.
The embedded ANO strings do not have this property, since they can terminate
on magnetic monopoles \cite{semilocal-review}. Superconducting semilocal strings have a
global current flowing through them,
whose conservation implies that they should either be infinitely
long or form closed loops. Such loops will in general be charged and will also have a
momentum circulating along them, and so that there is a possibility that they could
form quasi-stationary objects of vorton type. In fact there have been attempts to
construct such closed flux loop solutions in the SU(2) EAH model, although
the results obtained remain somewhat controversial for the time being
since the existence of such solutions, advocated in Ref.\cite{Niemi}, has not
been confirmed in the analysis of Refs.\ \cite{Ward}.

Maybe the most important open problem related to the new solutions
is their stability. Since for a fixed value of the current
the fundamental twisted solution have the least energy, we expect
them to be stable in the semilocal model.
However, a detailed stability analysis is necessary to make definite
conclusions.

Another interesting issue is to investigate the interactions between twisted vortices.
While type II ANO vortices (with $\beta>1$) always repel each other,
for our twisted vortices the situation seems to be more complicated.
Since we have superimposed vortex solutions for $n>1$,
we can ask  if it is energetically favorable for them to break up
into widely separated constituents with $n=1$.
The curves in Fig.\ref{fig-current}, where the ratio ${E}/(2\pi n)$ is plotted
against $\cal I$,  indicate that the interaction is repulsive up to a certain value
of $\cal I$, however, at least for $n=2$  it
becomes attractive for large enough currents.
For the $n=2$ vortex a clear crossover takes place around ${\mathcal I}/2\pi\approx0.51$.
This suggests that the $n=2$ superimposed vortex for large enough currents
becomes stable with respect to breakup into $n=1$ vortices, unlike
the type II ANO vortices. If the interaction of separated
vortices has also an attractive phase this could have important
physical consequences.
Clearly a more detailed analysis of this issue is necessary, which is,
however beyond the scope of our paper.
It is also worth mentioning
that in a different two-component Ginzburg-Landau model
 a similar non-monotonic behaviour of $E/n$ has been reported in
\cite{Babaev1}.

Before finishing our discussion, we would like to comment on the fact that
the vortex charge $\mathcal{I}_0$,
momentum $P$, and angular momentum $J$
vanish in the rest frame,
but the current $\mathcal{I}_3$
does not. This can certainly be explained using the tensorial
lows of transformation of the corresponding densities, as was done above.
At the same time, it might be worth illustrating the same fact also
in the framework of a simple
kinematic model,
viewing the superconducting string as consisting of positive and
negative elementary charges of the same mass in stationary motion.
The positive charges spiral around the vortex
axis clockwise in the positive $z$ direction, while the negative charges
spiral with the same
speed but counterclockwise and in the negative $z$ direction.
The positive charges have thus positive momentum and angular momentum, while
the negative
charges have negative momentum and angular momentum, but
the current is always oriented clockwise in the positive $z$ direction.
In the vortex rest frame
the volume densities of the positive and negative charges are thus identical, implying
that the total charge density, momentum, and angular momentum vanish, while
the total current does not. If the vortex is boosted in the
positive $z$ direction, say, then, due to the Lorentz contraction, the volume
density of positive charges will become higher as compared to the density
of negative charges. As a result, the total charge, momentum and angular momentum
will no longer be zero.

\end{document}